\documentclass[fleqn,usenatbib]{mnras}

\usepackage{newtxtext,newtxmath}

\usepackage[T1]{fontenc}
\usepackage{CJK}
\usepackage[normalem]{ulem}

\newcommand{\de}{\mathrm{d}}

\newcommand{\mathhangeulga}{\text{\it\begin{CJK}{UTF8}{mj}가\end{CJK}}}
\newcommand{\mathhangeulna}{\text{\it\begin{CJK}{UTF8}{mj}나\end{CJK}}}
\newcommand{\mathhangeulda}{\text{\it\begin{CJK}{UTF8}{mj}다\end{CJK}}}

\usepackage{graphicx,xcolor}	
\usepackage{amsmath}	
\usepackage{bm,mathtools}

\title[A measurement of the ISW with RACS]{A measurement of the Integrated Sachs-Wolfe Effect with the Rapid ASKAP Continuum Survey}

\author[B. Bahr-Kalus, et al.]
{Benedict Bahr-Kalus,$^{1}$\thanks{E-mail: benedictbahrkalus@kasi.re.kr}
David Parkinson,$^{1,2}$\thanks{E-mail: davidparkinson@kasi.re.kr}
Jacobo Asorey,$^{3}$
Stefano Camera,$^{4,5,6}$
Catherine Hale,$^{7}$ \and Fei Qin (\begin{CJK}{UTF8}{mj}秦斐\end{CJK})$^{1}$
\\
$^1$ Korea Astronomy and Space Science Institute, Yuseong-gu, Daedeok-daero 776, Daejeon 34055, Korea\\
$^2$ University  of  Science  and  Technology,  Daejeon  34113,  Korea\\
$^3$ Centro de Investigaciones Energeticas, Medioambientales y Tecnologicas (CIEMAT), Av.\ Complutense, 40, 28040 Madrid, Spain\\ 
$^4$ Dipartimento di Fisica, Universit\`a degli Studi di Torino, Via P.\ Giuria 1, 10125 Torino, Italy\\
$^5$ INFN -- Istituto Nazionale di Fisica Nucleare, Sezione di Torino, Via P.\ Giuria 1, 10125 Torino, Italy\\
$^6$ Department of Physics and Astronomy, University of the Western Cape, 7535 Cape Town, South Africa\\
$^7$ School of Physics and Astronomy, University of Edinburgh, Royal Observatory, Blackford Hill, Edinburgh EH9 3HJ, UK}

\date{Accepted XXX. Received YYY; in original form ZZZ}

\pubyear{2022}

\begin{document}
\label{firstpage}
\pagerange{\pageref{firstpage}--\pageref{lastpage}}
\maketitle

\begin{abstract}
The evolution of the gravitational potentials on large scales due to the accelerated  expansion of the Universe is an important and independent probe of dark energy, known as the integrated Sachs-Wolfe (ISW) effect. We measure this ISW effect through cross-correlating the cosmic microwave background maps from the \textit{Planck} satellite with a radio continuum galaxy distribution map from the recent Rapid ASKAP Continuum Survey (RACS). We detect a positive cross-correlation at $\sim 2.8\,\sigma$ relative to the null hypothesis of no correlation. We parameterise the strength of the ISW effect through an amplitude parameter and find the constraints to be $A_{\mathrm{ISW}} = 0.94^{+0.42}_{-0.41}$, which is consistent with the prediction of an accelerating universe within the current concordance cosmological model, $\Lambda$CDM. The credible interval on this parameter is independent of the different bias models and redshift distributions that were considered when marginalising over the nuisance parameters. We also detect a power excess in the galaxy auto-correlation angular power spectrum on large scales ($\ell \leq 40$), and investigate possible systematic causes.
\end{abstract}




\section{Introduction}

The mysterious acceleration of the expansion of the Universe, generated by the so-named dark energy, is now an established part of the concordance cosmological model, $\Lambda$CDM. The observational evidence comes not only from standard-candle and standard-ruler measurements of the expansion history but also from observations of the large-scale structure of matter and the distribution of the gravitational potential. 

An accelerating expansion will act against gravitational in-fall, slowing the accretion rate and decreasing the growth rate of cosmic structures. These structures, and their evolution in time, are observed through tracer particles. For high-redshift observations, the tracers are the photons emitted at the surface of last scattering, which form the cosmic microwave background (CMB) and trace the density fluctuations at recombination through the anisotropies in the intensity (i.e.\ temperature) and polarisation maps. In the CMB temperature power spectrum, the large-scale anisotropy is generated by the Sachs-Wolfe effect \citep[SW;][]{1967ApJ...147...73S} at last scattering, a gravitational redshift effect from photons climbing (or falling) out of the gravitational potential to enter the homogeneous universe. 

There is also a secondary effect generated long after recombination, caused by further evolution of the gravitational potentials, which is known as the integrated SW (ISW) effect. This late-time evolution of the potentials is driven by the accelerating universe, as the redshifting and blueshifting of photons moving into and out of density fields no longer exactly balances, but leaves some energy imprint in the photon frequencies. This process is an independent probe of the dark energy but it is difficult to see the effect on the CMB power spectrum alone. However, since the photon energies become correlated with the matter distribution at late times, the effect can be seen in the correlation between these two tracer fields \citep{PhysRevLett.76.575}.

The ISW effect was first detected in cross-correlation using NVSS 1.4 GHz radio catalogue \citep{1998AJ....115.1693C} and the HEAO1 A2 full-sky hard X-ray map \citep{BOLDT1987215} for large-scale structure tracers, and all-sky CMB map from the Wilkinson Microwave Anisotropy Probe \citep[\textit{WMAP};][]{Bennett_2003}, with a combined detection significance of $2.5\,\sigma$ \citep{Boughn2004,Boughn_2005}. The statistical significance of this ISW detection with the NVSS sample was reassessed by \citet{2008MNRAS.386.2161R}, which examined the consistency of the modelled bias-weighted redshift distribution with the data, giving an adjusted $3\,\sigma$ detection.

The ISW effect has also been detected using optical and infrared galaxies, cross-correlating \textit{WMAP} with galaxy samples extracted from the Automated Plate Measurement survey \citep*[APM; ][]{Fosalba:2003ge,Fosalba:2003iy}, the Sloan Digital Sky Survey \citep[SDSS;][]{Cabre:2006qm}, the 2MASS sample \citep{2011A&A...534A..51D}, and WISE galaxies \citep{PhysRevD.91.083533}. Recent work has updated the CMB maps from the \textit{Planck} mission \citep{Planck_ISW}, and detected the ISW in cross-correlation at $4\,\sigma$, again using the NVSS catalogue, as well as optical galaxies from the SDSS,  infrared galaxies from the WISE survey, and the \textit{Planck} 2015 convergence lensing map, as the low-redshift mass tracers. Most recently it has been detected using the DR8 galaxy catalogue of the DESI Legacy imaging surveys \citep{2021MNRAS.500.3838D}, using a ``low-density position'' filter, with a significance of $3.2\,\sigma$.

However, all of these detections are at a relatively low significance and have not added so much to the total constraining power of a cosmological data compilation. The next generation of surveys, like those proposed for the Australian Square Kilometre Array Pathfinder \citep[ASKAP;][]{Johnston:2008hp,hotan:2021} and the SKA Observatory,\footnote{\url{https://www.skatelescope.org}} will detect objects down to a lower surface brightness, and this increase in number counts should in turn increase the significance of the ISW detection, as well as the utility of the measurement. As the number count is increased, the sample can be split into redshift bins, which would make such a sample more sensitive to the long-wavelength radial power that generates the signal, and allow it to be used for more than a simple detection of the dark energy \citep[see][]{2012MNRAS.427.2079C,Ballardini:2018cho}. In \citet{Raccanelli:2014kga}, the authors forecast the effectiveness of such a future sample in determining the amplitude of the non-Gaussian contribution to the primordial density fluctuation. They found it to increase the effectiveness of \textit{Planck} for such, and be competitive with an all-sky optical survey such as that proposed for the \textit{Euclid} satellite \citep[see also][]{Alonso:2015sfa,Camera:2015yqa}. Similar forecasts have been made for the effectiveness of measuring primordial non-Gaussianity using the multi-tracer technique \citep{2014PhRvD..90h3520Y,2017MNRAS.466.2780F,2020MNRAS.492.1513G}, showing a predicted improvement over the constraints from \textit{Planck} alone.

In this work, we present our analysis of the cross-correlation of the CMB maps from the \textit{Planck} mission with a new radio continuum data set from the ``band 1'' sample \citep{RACS_data} of the first data release of the Rapid ASKAP Continuum Survey \citep[RACS;][]{McConnell2020}. RACS is a large-area radio continuum survey, covering the sky south of $+41^\circ$ declination. It is comparable to NVSS in depth, size of catalogue and area covered. It is different from NVSS in two key aspects. Firstly, it covers southern regions un-surveyed by NVSS. Secondly, whilst observations and reobservations were taken over 2019-2020, the  total on-source time was only a few weeks \citep[see ][]{McConnell2020}. RACS demonstrates the impressive survey power of ASKAP and provides an opportunity to test the cosmological analysis methods for the Evolutionary Map of the Universe (EMU) survey \citep{EMU,Norris:2021yqm}.

In \autoref{sec:theory}, we review the theoretical basis for the ISW effect. In \autoref{sec:data}, we describe our data sample and the methods and tools we use to analyse it. In \autoref{sec:results}, we give our results, and in \autoref{sec:summary} we summarise our findings.

\section{Theory}
\label{sec:theory}

The angular power spectrum of a set of tracers $X$ (e.g.\ galaxies, or photons) can be measured from the over-density field $\delta_X(\btheta)$ (where $\btheta$ is a particular direction on the sky)  
\begin{equation}
\label{eqn:alm}
a_{\ell m}^X = \int \de^2\btheta\;Y^*_{\ell m}\, \delta_X(\btheta)\;.
\end{equation}
Note that this is valid for a continuous density field. For a discrete density field, the integral is replaced with a sum.

Assuming an isotropic universe, we get the power spectrum from the auto-correlation
\begin{equation}
\label{eqn:angpowalm}
\left\langle a^X_{\ell m}\,a^{*X}_{\ell' m'}\right\rangle = \delta^{\rm K}_{\ell \ell'}\,\delta^{\rm K}_{mm'}\,C^{XX}_{\ell}\;,
\end{equation}
where $\delta^{\rm K}$ is the Kronecker symbol, and $C^{XX}_{\ell}$ is the angular auto-power spectrum of a map of tracer $X$ with itself. We can make a prediction for the angular power spectrum of a particular tracer using the three-dimensional power spectrum $P(k)$. Here the power needs to be averaged or `smoothed' in the radial direction, and the theoretical prediction is given by 
\begin{equation}
    C_\ell^{XX} = \frac2\pi\,\int\de k\;k^2\,P(k)\,\left[ W_\ell^X(k)\right]^2\;,
    \label{eq:Cl}
\end{equation}
where
$W_\ell(k)$ is the window function for the tracer $X$.

The galaxy window function (at linear order) is given by \cite[e.g.][]{2008PhRvD..77l3520G,2008MNRAS.386.2161R} \
\begin{equation}
  W_\ell^\mathrm{g}(k) = \int\de z\; n(z)\,b(z)\,D(z)\,j_\ell[k\,r(z)]\;,
  \label{eqn:galwindow}
\end{equation}
where $n(z)\,\de z$ is the source distribution per steradian with redshift $z$ within $\de z$ (brighter than some survey magnitude or flux limit), $b(z)$ is the linear bias factor relating tracer over-density to matter over-density, $D(z)$ is the growth factor of density perturbations, $j_\ell$ is the spherical Bessel function of order $\ell$, and $r(z)=\eta_0-\eta(z)$ is the radial comoving distance to redshift $z$, with $\eta(z)$ the conformal time coordinate at redshift $z$.

The cross-correlation power spectrum between a density field of large-scale structure tracers at low-redshift and the CMB temperature fluctuations is given by
\begin{align}
    C_\ell^{\rm gT} \coloneqq&\; \left\langle a_{\ell m}^{\mathrm{g}}\,a_{\ell m}^{*\mathrm{T}}\right\rangle\\
 =&\frac2\pi\,\int\de k\;k^2\,P(k)\,W_\ell^{\rm g}(k)\,W_\ell^{\rm T}(k)
    \label{eq:cl_isw}
\end{align}
where we now have two different window functions: $W_{\ell}^{\mathrm{g}}(k)$ for the large-scale structure tracer at low-redshift, and $W_{\ell}^{\mathrm{T}}(k)$ for the CMB photons. The window function for the CMB photons has a different structure to \autoref{eqn:galwindow}, as it is the power that is induced in the CMB temperature from the ISW effect, given by the equation
\begin{equation}
  \label{eqn:ISW}
  \left(\frac{\Delta T}{T}\right)_{\rm ISW}\hspace{-10pt}(\bm{x}_0,\btheta) = 2 \int_{\eta_{\rm dec}}^{\eta_0}\de\eta\; \dot\Upsilon[x_0-\btheta(\eta-\eta_0),\eta]\;,
\end{equation}
where $\dot\Upsilon$ is the time-derivative of the lensing potential (i.e.\ the Weyl potential) $\Upsilon = (\Phi+\Psi)/2$, with respect to conformal time $\eta$. Here, $\bm x_0$ is the observer's position (the photon position at time $\eta_0$), and $\btheta$ is the photon position at some general time.

Assuming no anisotropic stress, i.e.\ $\Phi = \Psi = \Upsilon$, the lensing potential obeys the field equation
\begin{equation}
    \ddot \Upsilon + 3\,\mathcal{H}\,\dot\Upsilon + (2\,\dot{\mathcal{H}} + \mathcal{H}^2) = 4\,\pi\,G\,a^2\,(\delta p),
    \label{eq:Upsilon_FE}
\end{equation}
where $\mathcal{H}$ is the conformal-time Hubble-Lema\^itre rate and $(\delta p)$ denotes the 1st-order perturbation on top of homogeneous and isotropic pressure.
Solving the Friedmann equations for a matter dominated universe, one gets $\mathcal{H} = 2/\eta$ and, thus, $2\,\dot{\mathcal{H}} + \mathcal{H}^2 = 0$. As, on cosmic scales, matter is a pressureless fluid, i.e.\ $(\delta p) = 0$, \autoref{eq:Upsilon_FE} simplifies to
\begin{equation}
    \ddot \Upsilon + \frac{6}{\eta}\,\dot\Upsilon = 0.
\end{equation}
The solution of this equation has the form
\begin{equation}
    \Upsilon = \aleph + \beth\,\eta^{-5}.
\end{equation}

Now, unless $\aleph$ is fine-tuned to be vanishingly small, we already have $\aleph \gg \beth\, \eta^{-5}$ at the epoch when the CMB photons are released and $\Upsilon$ is effectively constant during matter domination. Hence, as with \autoref{eqn:ISW}, the CMB photons retain the integrated history of the gravitational evolution of the Universe, a non-vanishing $(\Delta T/T)_{\rm ISW}$ proves that the Universe has undergone epochs where the cosmic fluid was not primarily composed of baryonic or dark matter ($\Omega_{\rm m} \neq 1$). In the concordance model of cosmology, these epochs are the radiation dominated epoch at early times when the CMB was released, and at late times, our current epoch, which is dominated by dark energy. As we are studying the ISW effect in cross-correlations between CMB anisotropies and the matter density field at relatively low redshifts $z \lesssim 5.2$, our analysis will establish evidence for or against the existence of dark energy.

To decrease the noise of the measured power spectra, we bin in multipole bins of width $\Delta\ell = 20$. We obtain the binned model power spectrum
\begin{equation}
    C_{\ell}^{XY, \mathrm{binned}} = \frac{\sum_{\ell^\prime \in \ell\text{-bin}}\,\ell^\prime\,(\ell^\prime + 1)\,C_{\ell^\prime}^{XY}}{\sum_{\ell^\prime \in \ell\text{-bin}}\,\ell^\prime\,(\ell^\prime + 1)}
    \label{eq:cl_binning}
\end{equation}
as the weighted average of the unbinned $C_{\ell}^{XY}$, where the $\ell^\prime(\ell^\prime + 1)$-weights are proportional to the variance, in turn, minimising the variance on $C_{\ell}^{XY, \mathrm{binned}}$ compared to a $(2\ell^\prime + 1)$-weighting scheme that corresponds to the number of modes entering each multipole $\ell$. Since we apply the same weights to both the data and the models used to infer covariances and the significance of our findings, our conclusions are unaffected by the choice of weighting scheme.

Finally, a computation of the theoretical power can be increased in speed by making the Limber approximation \citep{1953ApJ...117..134L}
\begin{equation}
    j_\ell[k\,r(z)]\xrightarrow{\ell\gg1} \sqrt{\frac{\pi}{2\ell + 1}}\,\delta^\mathrm{D}\left(\ell+\frac{1}{2} - k\,r(z)\right),
\end{equation}
which approximates the full window function calculation and convolution to a simple distance integral, with $\delta^D$ the Dirac distribution. This approximation breaks down when we integrate over more angular than radial modes. Hence, applying the Limber approximation at multipoles $\ell$ below some $\ell_\mathrm{min}$ can lead to catastrophic biases in the cosmological parameters of interest, e.g.\ as illustrated by \citet{Bernal:2020pwq} and proven by \citet{2022MNRAS.510.1964M} with a realistic analysis of a synthetic data set. However, in this instance, we are saved by not being able to locate radio continuum galaxies in redshift, thus, radial modes dominate even low multipoles and, hence, $\ell_\mathrm{min}$ becomes a function of the width of the redshift bin. \citet*{Tanidis:2019fdh} have estimated $\ell_\mathrm{min} = 2$ for a one-redshift-bin EMU-like survey. We are going to confirm the validity of the Limber approximation for our purposes in the following section before using it in cosmological analyses.

\section{Data Analysis}
\label{sec:data}

In this section, we describe the input data catalogues that we use, as well as the angular selection functions and the estimators that we employ to measure the angular power spectrum.

\subsection{Radio Data}
\label{sec:catalogues}

The radio data used in this work is from RACS \citep{McConnell2020, RACS_data}, an ASKAP survey that aims to observe the entire Southern sky ($\mathrm{Dec} \lesssim +41^{\circ}$) using a rapid survey strategy in three frequency bands over the $700-1800\, \mathrm{MHz}$ range. Each frequency band will use a bandwidth of $288\, \mathrm{MHz}$ for the observations. The first such data release, \citet{McConnell2020}, comprises images covering the Southern sky at $\mathrm{Dec} \lesssim +41^{\circ}$ and centred at a frequency of $888\, \mathrm{MHz}$, using 15-minute on-source observations. This is the lowest frequency band that will be observed with RACS. As part of the associated data release with \citet{McConnell2020}, images and catalogues were released covering 903 pointings, each with varying angular resolution across the sky.

For this work, it is essential to have a single catalogue across the sky without any duplication. Therefore we used the catalogue released within the second RACS paper \citep{RACS_data}, which we shall briefly discuss. In \cite{RACS_data}, the images of \cite{McConnell2020} were convolved to a common resolution of $25^{\prime\prime}$ and mosaicked together to produce a contiguous image across the majority of the sky covered by RACS. Convolving the image to a common resolution was essential to retain flux scale across the images before mosaicking. This resulted in 799 pointings which had sufficient resolution to be convolved to $25^{\prime\prime}$ and hence mosaicked together. The missing regions compared to \cite{McConnell2020} were concentrated in the Dec $ = +30^{\circ}$ to $+40^{\circ}$ regime and Dec $ = -90^{\circ}$ to $-80^{\circ}$. After mosaicking, sources were detected by running the source extraction software  \texttt{PyBDSF} \citep{pybdsf} over each of the 799 tiles using a $5\sigma$ criterion. The catalogues from the 799 tiles were then combined to avoid duplication, and to remove the Galactic plane, namely Galactic latitude between $-5^{\circ}$ and $+5^{\circ}$. The raw RACS over-density field from \citet{RACS_data} is mapped in \autoref{fig:racs_data}.

\subsubsection{Radio Data Weighting Function}
\label{sec:radiodataweightingfunction}
Despite the radio data catalogue from \cite{RACS_data} having uniform resolution across the sky, it is not uniformly sensitive across the images. This is due to a variety of factors: bright sources in the field affecting the neighbouring image, hour angle coverage differing with observations and the amount of overlap in mosaicking with neighbouring tiles. We, therefore, use the completeness simulations from \citet[][using resolved sources]{RACS_data} to determine the detection fraction of sources within each \texttt{HEALPix} bin.

\begin{figure}
	\centering
	\includegraphics[width=\columnwidth]{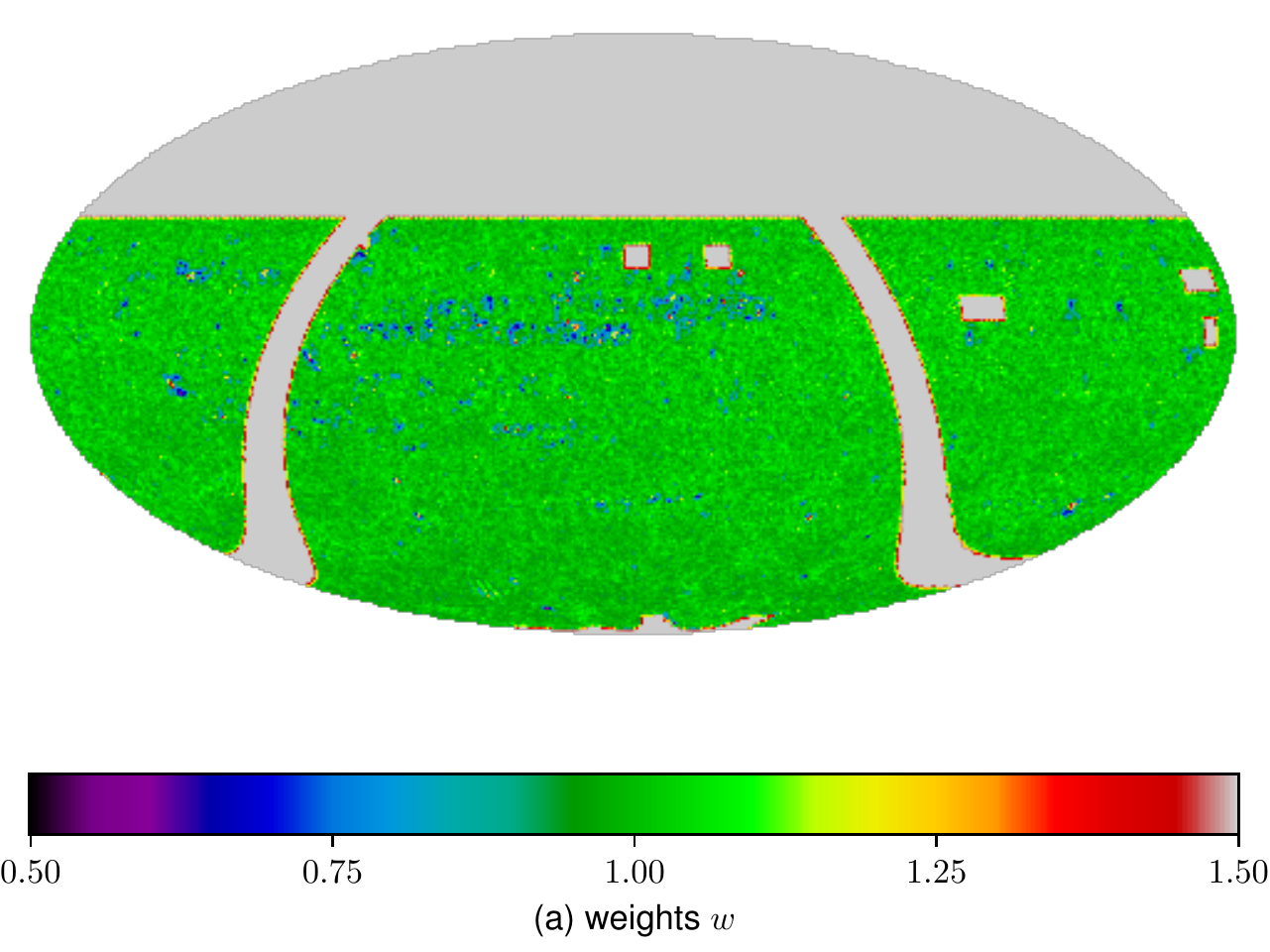}
	\includegraphics[width=\columnwidth]{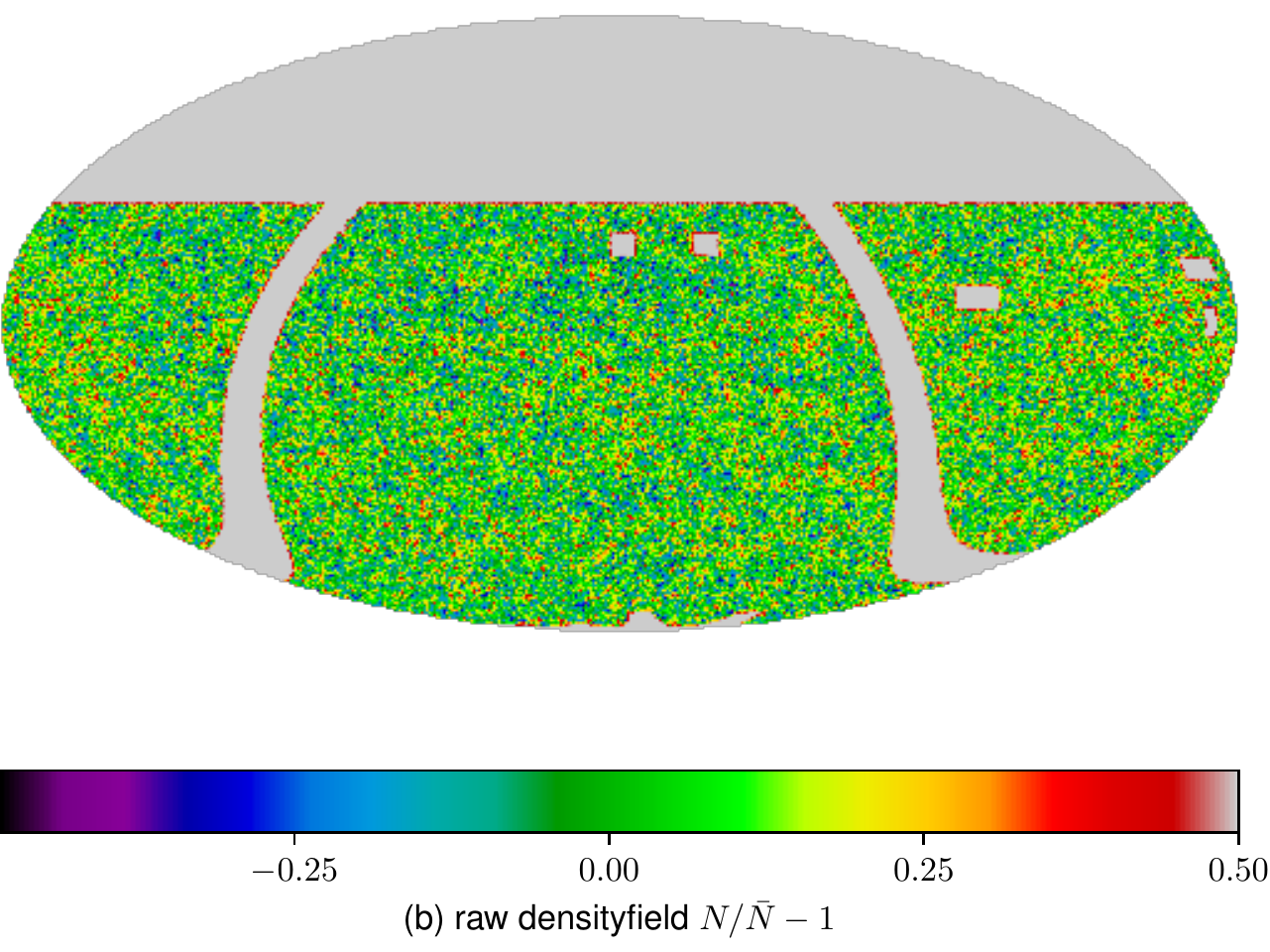}
	\includegraphics[width=\columnwidth]{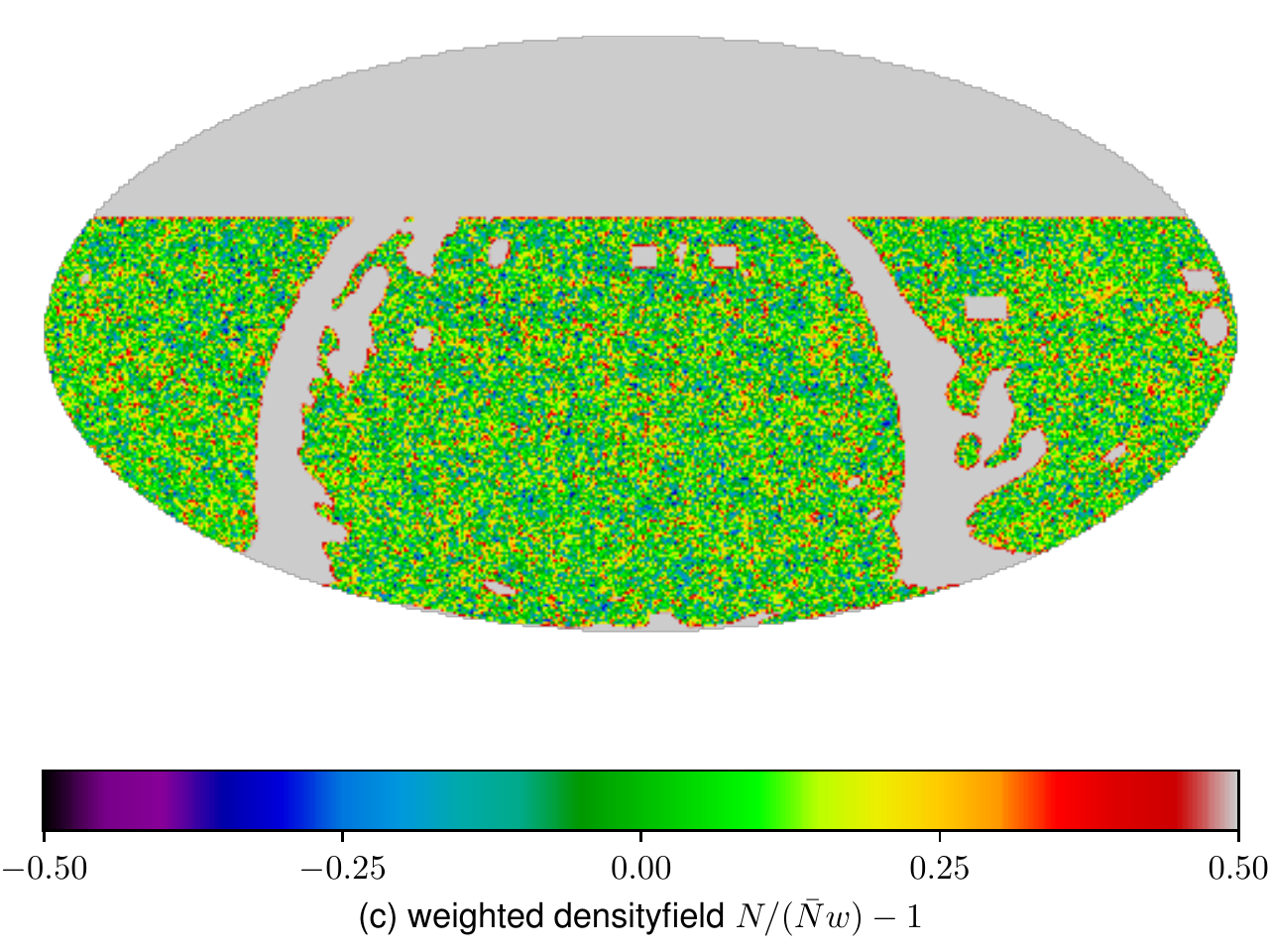}
    \caption{The distribution of the weights (\textit{top}), normalised number counts of objects above a flux density threshold of 4 mJy (\textit{middle}) and weighted over-density field (\textit{bottom}) of RACS radio continuum galaxies (still with 4 mJy flux density limit) on the sky after masking. All maps are in Mollweide projection and equatorial coordinates in astronomical orientation, i.e.\ showing East on the left-hand side.}
    \label{fig:racs_data}
\end{figure}

The simulations from \cite{RACS_data} use simulated sources from \cite{skads, 2010MNRAS.405..447W} and inject sources into the residual images and re-extract the sources using \texttt{PyBDSF}. These simulations use 5 million random sources across  Dec $ = -85^{\circ}$ to $+30^{\circ}$ and each simulation is repeated 10 times. We combine all the recovered sources (which have a output ``measured'' flux that would have resulted in a $5\sigma$ detection)  within a \texttt{HEALPix} bin and compare this to the number of sources within the \texttt{HEALPix} bin that were injected to determine the weight within a given \texttt{HEALPix} bin. A map of the radio data weights is shown in the top panel of \autoref{fig:racs_data}.

\subsubsection{The Radio Dipole}
\label{sec:dipole}

The angular two-point statistics of the NVSS catalogue initially showed an excess at large scales that could have been interpreted as the signal due to the scale dependent bias effect due to a non-Gaussian distribution of the primordial density field \citep{Xia:2010yu}. This excess signal has disappeared after \citet{Chen:2015wga} reanalysed the NVSS catalogue using a new mask taking sidelobe effects of bright sources, the Galactic foreground and the radio dipole signal into account. The radio dipole is believed to result from our peculiar motion whose velocity vector (in natural units) is written as $\bm v_\mathrm{pec}$.

While the first two issues are addressed by the weighting function described in Sec.~\ref{sec:radiodataweightingfunction}, the latter modulates the observed density field \citep{Ellis:1984}. Assuming that the flux density $S$ at a given frequency $\nu$ and the number count $\bar N(> S)$ are both given by power laws with, respectively, spectral indices $\alpha$ and $x$, i.e.\
\begin{align}
    S&\propto \nu^{-\alpha}\;,\\
    \bar N(> S) &\propto S^{-x}\;,
\end{align}
the observed density field $\delta_\mathrm{g}^\mathrm{obs}$ in direction $\btheta$ is given by \citep[e.g.][]{Bengaly:2018ykb}
\begin{equation}
    \delta_\mathrm{g}^\mathrm{obs} = \delta_\mathrm{g}^\mathrm{rest} + \left[2 + x\,(1 + \alpha)\right]\,\btheta\cdot \bm v_\mathrm{pec}\;,
    \label{eq:deltarest}
\end{equation}
where $\delta_\mathrm{g}^\mathrm{rest}$ represents the over-density field in the rest frame where the galaxy distribution is statistically isotropic.

We coincidentally estimate $\alpha = x = 0.76$ from SKADS (which we adopt from here on), as well as $\alpha = 0.82$ and $x = 0.90$ from T-RECS. While these simulation-based estimate might not provide us with the most accurate measurement of $\alpha$ and $x$ \citep[for instance, SKADS underestimates source counts at faint flux densities, see e.g.][]{Smolcic:2016sma, Norris:2021yqm, Gurkan2022} they are consistent with observations at higher flux densities and with that of \citet[][]{RACS_data} above $\sim$2 mJy. Measurements of $\alpha$ are commonly measured from radio surveys and assumed in studies within the literature to be $\sim0.7-0.8$ \citep[see e.g.][]{Smolcic:2016sma, deGasperin2018, Norris:2021yqm}, though we note that \cite{RACS_data} found slightly larger/smaller values dependent on the frequency being compared to. The scatter between the SKADS and T-RECS results is also dwarved by the scatter among different measurements of the amplitude of $\bm v_\mathrm{pec}$ from radio surveys, such as \citet{Blake:2002gx,Singal:2011dy,Gibelyou:2012ri,Rubart:2013tx,Kothari:2013gya,Tiwari:2013ima,Tiwari:2015tba}. \citet*{Siewert:2020krp} even find an apparent frequency dependence of the radio dipole amplitude. Given this uncertainty in the amplitude and the fact that all of them agree in direction with the CMB dipole, a more natural assumption of $\bm v_\mathrm{pec}$ when subtracting the second term of \autoref{eq:deltarest} is the CMB dipole measured by \textit{Planck} \citep{Planck:2018nkj}. By doing so, we are also consistent with the CMB data that we describe in the next subsection, from which the CMB dipole has been subtracted.

\subsection{CMB Data}
\label{sec:CMBdata}

We make use of the third release SMICA \textit{Planck} Legacy Map \citep{Planck_maps}. SMICA \citep{Delabrouille:2002kz,Cardoso:2008ISTSP...2..735C} stands for Spectal Matching Independent Component Analysis and is one of the four component separation methods used by the \textit{Planck} Collaboration. The SMICA data model
\begin{equation}
    \mathbfss{R}_\ell = {\mathbfss a\,\mathbfss a^\dagger}\,C_\ell^\mathrm{TT} + {\mathbfss A\,\bm P_\ell\,\mathbfss A^\dagger} + {\bm N}_\ell
\end{equation}
is a superposition of the true CMB signal (expressed in terms of the matrix $\mathbfss a$ composed of $a_{\ell m}^\mathrm{T}$ in each frequency band and their frequency-independent auto-power spectrum $C_\ell^\mathrm{TT}$), the noise spectrum ${\bm N}_\ell$ and foreground signals ${\mathbfss A\,\bm P_\ell\,\mathbfss A^\dagger}$. The foreground signals are expressed in terms of a small number of templates with arbitrary frequency spectra, arbitrary power spectra and arbitrary component correlations. These are fitted to the auto- and cross-power spectra of \textit{Planck} maps ${\bm x}_{\ell m}$ in its nine frequency channels. The final SMICA map,
\begin{equation}
    \hat {\bm s}_{\ell m} = {\mathbfss w^\dagger_\ell\,\bm  x}_{\ell m}\;,
\end{equation}
is then obtained by fitting weights (note that these are unrelated to the weights in \autoref{eq:cl_binning})
\begin{equation}
    {\mathbfss w}_\ell = \frac{\mathbfss{R}_\ell^{-1}\,\mathbfss a}{\mathbfss a^\dagger\,  \mathbfss{R}_\ell^{-1}\,\mathbfss a}
\end{equation}
that minimise the discrepancy between the frequency channel map auto- and cross-power spectra, i.e.\
\begin{equation}
    \widehat{\mathbfss w}_\ell = \arg\min_{{\mathbfss w}_\ell}\sum_\ell\; \left(\sum_m\; {\bm x}_{\ell m}\,{\bm x}^\dagger_{\ell m}\,{\mathbfss{R}_\ell} + (2\,\ell + 1)\ln\det{\mathbfss{R}_\ell}\right).
\end{equation}

The fit is done in three steps:
\begin{enumerate}
    \item Only the CMB power spectrum $C_\ell$ and $\mathbfss a$ are fitted on a clean patch of the sky;
    \item All other parameters are fitted on a large patch of the sky while keeping $\mathbfss a$ fixed at the best-fitting value of the previous step;
    \item $\mathbfss a$ and $\mathbfss A$ are fixed to their previously found values while the power spectra $C_\ell$ and $\bm P_\ell$ are fitted.
\end{enumerate}
SMICA is the foregound component separation method that has performed best in a \textit{Planck} foreground-cleaning mock challenge \citep{Planck:2013fzg}. However, we have found that the choice of component separation method has no significant impact on the galaxy-temperature cross-correlation, and thus, on the ISW signal. 

The temperature map can be retrieved as the \texttt{I\_STOKES} column from the FITS file downloadable from the digital object identifier given in the reference of \citet{Planck_maps}. We rotate and downgrade the resolution of the \textit{Planck} map from its initial $N_\mathrm{side} = 2048$ in galactic coordinates to match RACS's $N_\mathrm{side} = 128$ in equatorial coordinates. We perform the same transformations to the temperature confidence mask given in the \texttt{TMASK} column and we cut out pixels from the RACS map where the value of the temperature confidence is less than $0.5$. Equally, we mask out CMB pixels that are also masked out by the RACS mask. We show the binary mask outlining the quality cuts imposed on the RACS and \textit{Planck} data in grey in the bottom panel of \autoref{fig:racs_data}.

\subsection{Estimating the Angular Power Spectra}
\label{sec:estimators}

The estimation of the spherical harmonic amplitudes, and the angular power spectrum, as given in \autoref{eqn:alm} and \ref{eqn:angpowalm}, assumes that the full-sky is available. For a cut-sky, as we have with both the CMB and radio continuum data, we need to apply an angular selection function (as described in sections \ref{sec:radiodataweightingfunction} and \ref{sec:CMBdata}) and estimate from only those regions that are visible. This leads to measured amplitudes $\tilde{a}_{\ell m}$'s that are different from the true values, and a pseudo angular power spectrum $\tilde{C}_\ell$, as computed by the MASTER algorithm \citep{2002ApJ...567....2H}. The advantage of the MASTER algorithm is that the measured $\tilde{C}_\ell$ can then be directly compared to the theoretical prediction. In this work, we use the python implementation of the algorithm, \texttt{NaMaster} \citep{2019MNRAS.484.4127A}.

Following the approach of \citet{2021MNRAS.502..876A}, we first generate a map of the radio continuum over-density field, which we do by combining the galaxy number count map $N(\btheta)$ with the radio data weighting function map $w(\btheta)$ from \autoref{sec:catalogues}, using the equation
\begin{equation}
    \delta_\mathrm{g}(\btheta\,) = \frac{N(\btheta)}{\bar{N}\,w(\btheta)} -1\;,
\end{equation}
where $\btheta$ is a particular direction (or \texttt{HEALPix} pixel) on the sky and $\bar{N}$ is the average weighted number of galaxies per \texttt{HEALPix} cell. To construct the over-density map, we cut all those pixels $\btheta$ that have weights $w(\btheta) < 0.5$, to prevent a bias. However, these are only a very small number that still lie inside the region selected in the angular window. The over-density field is shown in \autoref{fig:racs_data}.

As galaxies are discrete objects sampling the continuous density field, the pseudo galaxy auto-power spectrum $\tilde{C}_\ell^\mathrm{gg}$ will disagree with the model power spectrum $C_\ell^\mathrm{gg}$ by a constant shot-noise term $N_\mathrm{shot}$. Na\"ively, one can think of the galaxies being drawn from the matter field as a Poisson point process. In spite of that, some galaxies appear as multiple sources in a radio catalogue, whereas in other instances, multiple sources may not be identified as such by the source finder. Consequently, the shot noise level can deviate from its Poisson prediction. The source finding can be approximated as a supplementary Poisson sampling from the already Poisson sampled galaxy number count, resulting in a so-called compound Poisson distributed sample \citep{Siewert:2019poc}. In any case, the compound Poisson distribution also predicts a scale-independent shot-noise power spectrum and, instead of modelling it, we fit a constant $\hat N_\mathrm{shot}$ that minimises $(\tilde{C}_\ell^\mathrm{gg} - \hat N_\mathrm{shot} - C_\ell^\mathrm{gg, fid})^2$ for the hereinafter defined fiducial power spectrum $C_\ell^\mathrm{gg, fid}$.

\subsection{Theoretical Predictions and Modelling}
\label{sec:modelling}

\subsubsection{Cosmological Parameters}
\label{sec:cosmoparams}

To model theoretically the power spectra that we want to compare our data against, we assume a flat, homogeneous and isotropic universe where the laws of gravity are expressed by the theory of general relativity. As we cannot faithfully measure all cosmological parameters from RACS alone, we fix the parameters listed in \autoref{tab:cosmo_params} at the reported values. These are for the most part the default values of the `Code for Anisotropies in the Microwave Background' \citep[CAMB;][]{Lewis:1999bs,Howlett:2012mh}, with the exception of $n_\mathrm{s}$ and $\tau$ which we take from \citet{Planck:2018vyg} for consistency with the \textit{Planck} 2018 maps \citep{Planck_maps}. Note that parameters that are not matched to \textit{Planck} 2018 are within 1 sigma from the \textit{Planck} 2018 best-fitting values.

To check the validity of the Limber approximation (cf. \autoref{sec:theory}), we evaluate \autoref{eq:Cl} twice, once with and once without making use of the Limber approximation, for the same fiducial cosmology, bias and redshift distribution models. Binning the result in multipole bins with width $\Delta\ell = 20$ (as we are going to do in our analyses), we find a bias of $\sim1\%$ in the $C_\ell^\mathrm{gg}$ prediction for the lowest $\ell$-bin and much smaller biases at smaller scales. We revisit this assumption later when we evaluate the likelihood of our data.

As discussed in \autoref{sec:theory}, the existence of dark energy causes a correlation between the CMB temperature map and the distribution of matter due to the late-time ISW effect. On the other hand, in a Universe without significant dark energy that would be dominated by matter until our present epoch, we should not measure a notable cross-correlation between the two fields. We, therefore, introduce a phenomenological parameter $A_\mathrm{ISW}$, such that
\begin{equation}
    C_\ell^\mathrm{gT} = A_\mathrm{ISW}\,C_\ell^\mathrm{gT, fid},
\end{equation}
where $C_\ell^\mathrm{gT, fid}$ is the galaxy-temperature cross-power spectrum computed for the fiducial parameters listed in \autoref{tab:cosmo_params}. In this way, if we measure an $A_\mathrm{ISW}$ that is consistent with zero, we have not detected the ISW effect and, thus, we have found no evidence for dark energy. Should $A_\mathrm{ISW}$, however, be consistent with unity, then our $\Lambda$CDM-based model of the galaxy-temperature cross-power spectrum is consistent with the data. If $A_\mathrm{ISW}>0$ but inconsistent with one, then we still have detected dark energy but we have to revisit our modelling assumptions.
\begin{table}
    \centering
    \caption{Fiducial cosmological parameters assumed throughout this paper.}
    \label{tab:cosmo_params}    \begin{tabular}{l|c|c}
        \hline 
        Parameter & Symbol & Value/Relationship \\\hline
        Hubble-Lema\^itre constant & $H_0$ & $67.5\;\mathrm{km/s/Mpc}$\\
        Reduced Hubble-Lema\^itre constant & $h$ & $H_0/(100\;\mathrm{km/s/Mpc})$\\
        Physical baryon density parameter & $\Omega_\mathrm{b}$ & $0.022/h^2$\\
        Cold dark matter density parameter & $\Omega_\mathrm{cdm}$ & $0.12/h^2$\\
        Total matter abundance & $\Omega_\mathrm{m}$ & $\Omega_\mathrm{cdm}+\Omega_\mathrm{b}$\\
        Dark energy density parameter & $\Omega_\Lambda$ & 
        $1-\Omega_\mathrm{m}$
        \\
        Reionisation optical depth & $\tau$ & 0.0544\\
        Amplitude of scalar fluctuations & $A_\mathrm{s}$ & $2\times 10^{-9}$\\
        Scalar spectral index & $n_\mathrm{s}$ & 0.965\\
        \hline
    \end{tabular}
\end{table}

\subsubsection{Number Count Model}

\begin{figure}
	\includegraphics[width=\columnwidth]{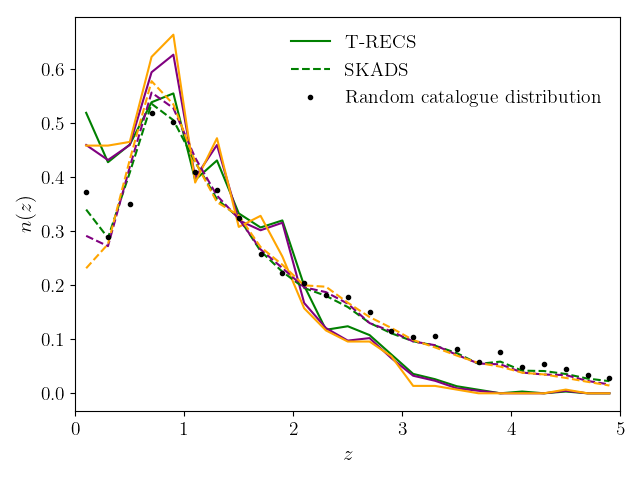}
    \caption{The distribution of radio continuum galaxies with redshift, as predicted by the SKADS (dashed) and T-RECS (solid line) mock radio continuum galaxy catalogues. The different colours correspond to $2$ (green), $3$ (purple) and $4\,\mathrm{mJy}$ (orange) flux density limits. The black dots are the distribution used in the generation of random catalogues, as described in \autoref{sec:radiodataweightingfunction}.}
    \label{fig:nz}
\end{figure}

To make accurate predictions for the angular power spectrum of a galaxy sample, the window function needs to be computed using some well-motivated estimate for the redshift distribution of galaxy number per steradian $n(z)$ and bias $b(z)$. For our sample of radio continuum galaxies (being observed at $\sim 1\,\mathrm{GHz}$), we make use of simulations to inform this redshift distribution. Two of the major existing extra-galactic radio simulations that are available to use are the European SKA Design Study (SKADS) Simulated Skies \citep{skads} and the Tiered Radio Extra-galactic Continuum Simulation \citep[T-RECS;][]{2019MNRAS.482....2B}. In \autoref{fig:nz}, we show the predicted $n(z)$ distribution for several different flux cuts from both the T-RECS and SKADS simulated catalogues, and the distribution used in the generation of random catalogues, as described in Sec.~\ref{sec:radiodataweightingfunction}. Although the predictions are very similar for 2, 3 and 4mJy, we assume a value of 4mJy for all theoretical predictions for the rest of the paper. This 4mJy reflects a region where, above this flux density limit, we believe the random weight maps appropriately account for incompleteness within the survey, as can be seen in the source counts corrections of \cite{RACS_data}. 

We see that both simulations make roughly similar predictions for the redshift distribution, peaking at around $z=1$ and slowly falling off at higher redshifts. However, the SKADS prediction has a larger high-redshift tail, with $90\%$ of galaxies lying below $z<3.6$. In contrast, the T-RECS galaxies are more localised to $z\sim1$, with $90\%$ lying below $z<3.1$. This will affect the power spectrum predictions, as the window function given in \autoref{eqn:galwindow} will average the radial fluctuations out over a larger range of $k$-values for SKADS than T-RECS, diluting the power and so leading to a lower amplitude for the same cosmology. We consider both $n(z)$ models in our analysis.

\subsubsection{Bias Model}
\label{sec:bias}

\begin{figure}
    \centering
    \includegraphics[width=\columnwidth]{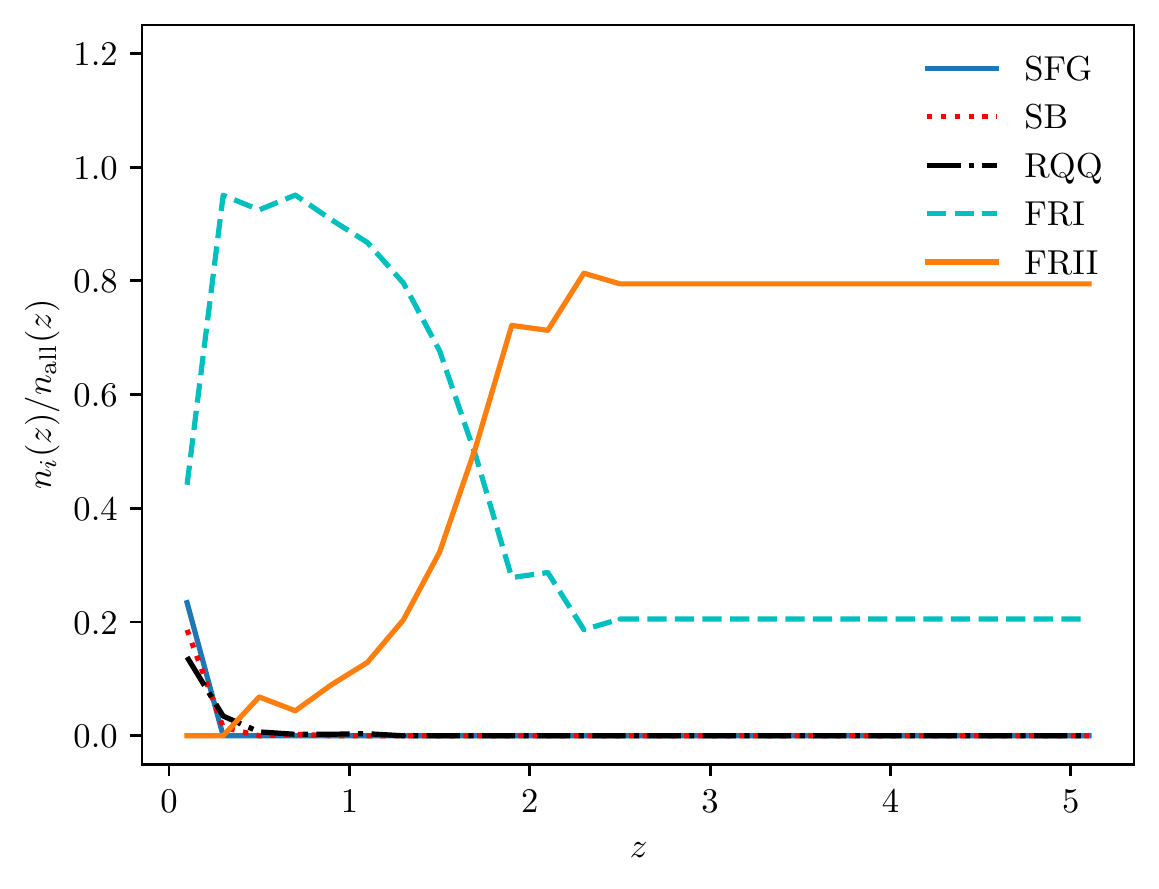}
    \caption{The predicted fraction of the observed number of star-forming galaxies (SFGs), star burst galaxies (SBs), radio-quiet quasars (RQQs), active galactic nuclei of type FRI or type FRII in the Fanaroff-Riley classification, and the total number of observed objects estimated from SKADS \citep{skads}. This prediction assumes a 4mJy flux cut.}
    \label{fig:pop_frac}
\end{figure}

Radio surveys are known to trace two galaxy populations: Active Galactic Nuclei (AGN) and Star Forming Galaxies (SFGs). 
The peak-background split model \citep{Bardeen:1985tr,Cole:1989vx} predicts the simple relationship 
\begin{equation}
    \delta_\mathrm{g} = b\,\delta_\mathrm{m}
\end{equation}
between the galaxy over-density field $\delta_\mathrm{g}$ and the matter over-density field $\delta_\mathrm{m}$. This in turn means 
\begin{align}
    C_\ell^\mathrm{gg} &= b^2\, C_\ell^\mathrm{mm}\;,\\
    C_\ell^\mathrm{gT} &= b\, C_\ell^\mathrm{mT}\;,
\end{align}
for the power spectra, under the assumption of a constant bias across redshift, i.e.\ $b(z)\equiv b$. As time progresses, more galaxies have the chance to form within haloes and evolve, thus the galaxy bias is in general a redshift dependent quantity.

For a combined sample, where the individual species of galaxies are not separated when the clustering is measured, the angular correlation function and power spectra are only sensitive to the total bias. For this total bias, we must combine the biases by weighting them with the individual number counts $n_i(z)$ of each galaxy type population as in \citet{2014MNRAS.442.2511F} \citep[see also][]{Bernal:2018myq,2020MNRAS.492.1513G,Asorey:2021hgc}, namely
\begin{equation}
    \label{eqn:total_bias}
    b(z) = \frac{\sum_i\;{b_i(z)\,n_i(z)}}{n_\mathrm{all}(z)}\;,
\end{equation}
where $i$ corresponds to the different populations and $n_\mathrm{all}(z)$ is the whole sample redshift distribution. Then, we need some prescription for the biases of the individual populations.

At low redshifts, we have some good measurements of the bias values of each population \citep[e.g.][]{Magliocchetti:2016jfv,Hale:2017wub,Dolfi:2019wye}. However, at higher redshifts the bias is a large source of uncertainty, amplified by our ignorance of what ratio of the observed population is composed of what type of radio source. For RACS, we estimate that SFGs make a considerable fraction of objects only at $z \sim 0.2$ (cf. \autoref{fig:pop_frac}).\footnote{Note that SKADS further subdivides this population into starburst galaxies (SBs) and true SFGs.} At higher redshifts, the RACS catalogue is dominated by AGNs. Up until redshift $z \sim 1.8$, most of them fall into the first Fanaroff-Riley class (FRI). Above that redshift, FRIIs are the most important radio source.

The fitting that was done as part of the \citet{skads} analysis gave a parameterised form of this $b_i(z)$, and these bias models have been used extensively in forecasting the potential that radio continuum surveys have to probe cosmology \citep[see e.g.][]{2012MNRAS.427.2079C,2014MNRAS.442.2511F,Raccanelli:2014kga,Bernal:2018myq,Asorey:2021hgc}, and are described in detail there. In these models each population has a bias that evolves exponentially with redshift. \citet{skads} argue that this leads to excessively strong clustering at high redshifts and, therefore, propose a constant bias above a certain cut-off redshift.

Instead of using theoretical models for the bias, that are based on $N$-body simulations, we can parameterise our ignorance, and attempt to measure the bias directly from the data. Here we consider the following effective $b(z)$ parameterisations (which we shall compare with the fiducial bias from SKADS and T-RECS in \autoref{sec:bias_results}):
\begin{enumerate}
    \item As can be seen in \autoref{fig:pop_frac}, the RACS catalogue is expected to be composed mostly of AGNs. For both AGN FR subtypes, the SKADS bias model plateaus above $z > 1.5$. We therefore consider an exponential bias \begin{equation*}b(z) = \begin{cases} b(0)\exp\left(\dfrac{\mathrm{d}\ln b}{\mathrm{d}\ln z}\right) & \text{ for } z < 1.5 \\ b(0)\exp\left(\left.\dfrac{\mathrm{d}\ln b}{\mathrm{d}\ln z}\right\vert_{z=1.5}\right) & \text{ for } z \geq 1.5\end{cases}\;,\end{equation*} with an arbitrary redshift cap at $z = 1.5$, motivated by \citet{skads}.  \label{item:expcutbias}
    \item Since we find the redshift cut-off somewhat arbitrary, we also study an exponential bias $b(z) = b(0)\exp(\mathrm{d}\ln b/\mathrm{d}\ln z)$ that is still well motivated at the redshift range where we expect the bulk of our observed objects.\label{item:expbias}
    \item A linear bias $b(z) = b(0) + \mathrm{d}b/\mathrm{d}\ln z$ that allows for redshift evolution without excessive clustering at the high-redshift tail. \label{item:linbias}
    \item Lastly, a constant bias $b$ that has been assumed in forecasts at high redshift. \label{item:constbias}
\end{enumerate}

\begin{figure}
    \centering
    \includegraphics[width = \columnwidth]{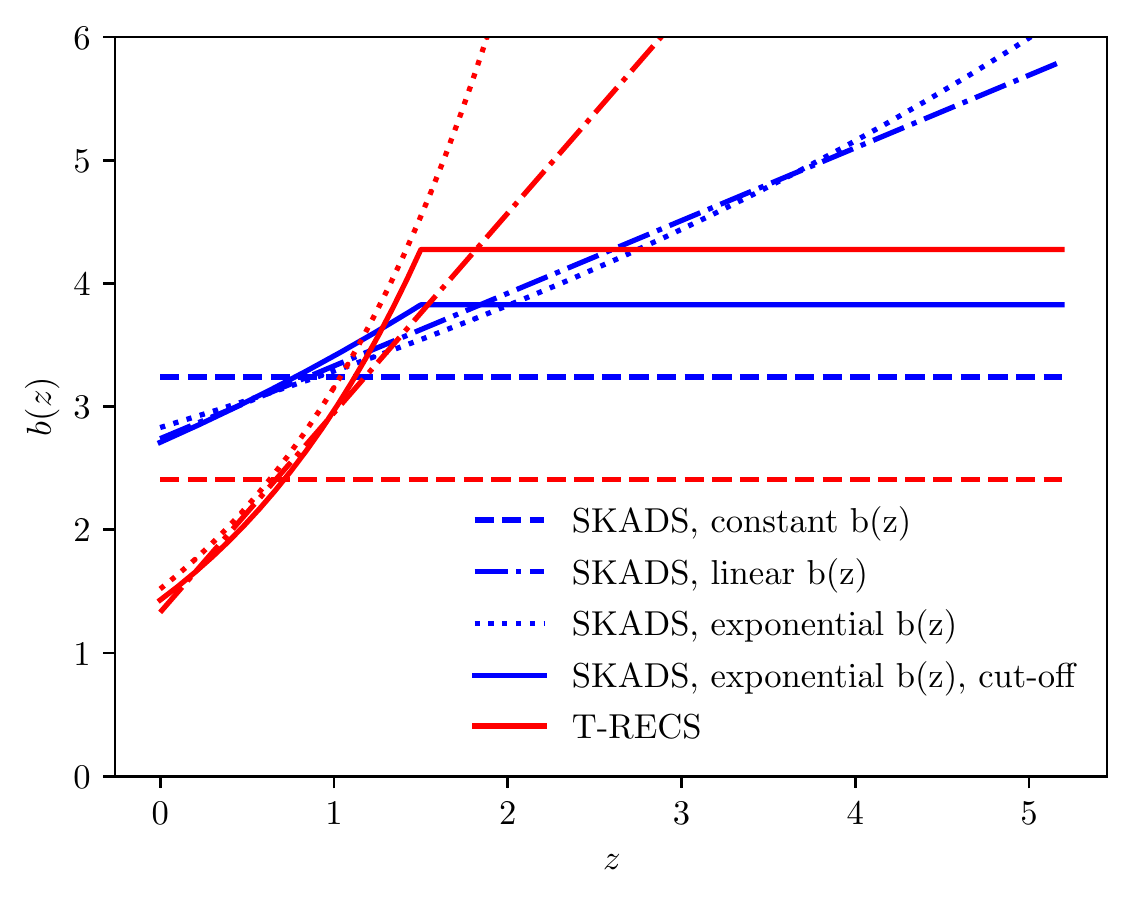}
    \caption{The bias parameterisations considered in this work evaluated at the best-fitting parameters for both SKADS- and T-RECS $n(z)$ distributions listed in \autoref{tab:b_evo_results}.}
    \label{fig:bias_models}
\end{figure}

These models are plotted in \autoref{fig:bias_models}. As galaxies only form in high density regions, $b(z)$ has to be positive. We therefore impose hard priors $b(0)>0$ in \ref{item:expcutbias}-\ref{item:linbias}, $\mathrm{d}b/\mathrm{d}z > b(0)/z_\mathrm{max}$ (with $z_\mathrm{max} = 5.2$ the assumed maximum redshift attainable by the survey) in \ref{item:linbias}, as well as $b>0$ in \ref{item:constbias}. Note that, since we keep $n(z)$ fixed in our analyses and $n(z)$ is degenerate with $b(z)$, our uncertainty on the bias $b(z)$ also effectively incorporates our uncertainty on $n(z)$.

\subsection{Mock Catalogues}
\label{sec:mocks}

We generate mock over-density fields to test our analysis pipeline as well as to estimate the statistical significance of our measurements. To do so, we use the publicly available Full-sky Lognormal Astro-fields Simulation Kit \citep*[\verb~flask~]{Xavier:2016elr} to draw Gaussian random fields for a given set of angular model power spectra $C_\ell^\mathrm{g}$ and $C_\ell^\mathrm{T}$. We also have the option to further condition each pair of mock- galaxy and CMB maps by defining a model $C_\ell^\mathrm{gT}$. For the CMB maps, we just mask out regions that are not observed in the data from the Gaussian realisations of the temperature maps. For the mock RACS source maps, we first let \verb~flask~ Poisson sample the number of mock sources
\begin{equation}
    n(\btheta)=\operatorname{Poisson}\left\lbrace\bar{n}\, w(\btheta)\,\left[1 + \delta_\mathrm{g}(\btheta)\right]\right\rbrace
\end{equation}
from the Gaussian density field realisations $\delta_\mathrm{g}(\btheta)$, where we choose the average number density $\bar n$ such that the total number of sources matches the number of sources in the data catalogue. We also apply the same completeness weights $w(\btheta)$ and mask as for the data. Finally, the mock source count and CMB maps are saved in the \texttt{HEALPix} format with $N_\mathrm{side}=128$.

\subsection{Covariance Matrices}
\label{sec:covariances}

We explore four different ways to estimate the covariances $\mathbfss{S}_{\ell\ell^\prime}^{WXYZ}$ between multipoles $\ell$ and $\ell^\prime$ and fields $W, X, Y$ and $Z$. In the general case where all fields can be different, we use an analytic estimate based on a fiducial power spectrum and a mixing matrix encompassing the effect of the survey mask. We use one internal method, i.e.\ estimating the covariance by resampling the data, as well as two external methods where we estimate the covariance matrices from mock realisations of the data. We can use the covariance matrices obtained in these different ways to validate them against each other. 

\subsubsection{Analytic Covariance}
\label{sec:analyticcov}
Given two maps $X(\bm\theta)$ and $Y(\bm\theta)$, measurements $\tilde C^{XY}_\ell$ and $\tilde C^{XY}_{\ell^\prime}$ of their harmonic-space cross-power spectrum at two different multipoles have covariance defined by
\begin{align}
    \mathbfss{S}_{\ell\ell^\prime}&\coloneqq{\rm Cov}\left[\tilde C^{XY}_\ell,\tilde C^{XY}_{\ell^\prime}\right] \nonumber\\ 
    &=\left\langle a^X_{\ell m}\,a^{\ast Y}_{\ell m}\,a^Y_{\ell^\prime m^\prime}\,a^{\ast X}_{\ell^\prime m^\prime}\right\rangle  -\left\langle a^X_{\ell m}\,a^{\ast Y}_{\ell m}\right\rangle\,\left\langle a^Y_{\ell^\prime m^\prime}\,a^{\ast X}_{\ell^\prime m^\prime}\right\rangle\;.
\end{align}
Under the hypothesis of Gaussianity, and using Wick's theorem to break up the four-point correlator into products of two-point correlators, we find
\begin{equation}
    \mathbfss{S}_{\ell\ell^\prime}=\frac{\tilde C^{XX}_\ell\,\tilde C^{YY}_\ell+\left(\tilde C^{XY}_\ell\right)^2}{(2\,\ell+1)\,\Delta\ell}\,\delta^{\rm K}_{\ell\ell^\prime}\;.
\end{equation}

In the case of partial sky coverage, a common approximation is to perform the rescaling $\mathbfss{S}_{\ell\ell^\prime}\to\mathbfss{S}_{\ell\ell^\prime}/f_{\rm sky}$, where $f_{\rm sky}$ is the fraction of the sky observed. For $f_{\rm sky}\lesssim1$, this approximation performs well and has the advantage of correctly accounting for the increase in the (co)variance of the measurements due to a more limited number of available modes. However, if $f_{\rm sky}$ is significantly smaller than unity, or if the survey mask is highly non-trivial, or if coverage and depths change across the sky, more refined methods are needed. As mentioned in \autoref{sec:estimators}, one of such methods is represented by pseudo-$C_\ell$'s, where the coupling between different multipoles induced by the partial sky coverage is encoded in the so-called coupling matrix---in turn, related to the power spectrum of the mask/weight map. Once this quantity is given, the \texttt{NaMaster} code allows for the evaluation of the masked covariance matrix.

\subsubsection{Jackknife Resampling}
\label{sec:jackknife}

Internal covariance matrix estimation methods have the advantage that they are independent of any cosmological model, the survey selection is naturally accounted for, and the contribution of hidden or unforeseen systematic errors is inherent in the uncertainties estimated by internal methods. On the other hand, they rely on the assumption that the data is an accurate representation of the distribution of measurements. Sampling fluctuations known in the cosmology literature as \textit{cosmic variance} are therefore not included in internal covariance matrix evaluations \citep[see e.g.][]{Norberg:2008tg}. 

We make use of the `delete one'-jacknife method proposed by \citet{Shao}. We draw $N_\mathrm{sub}$ subsamples of non-adjacent non-zero \texttt{HEALPix} cells without replacement, i.e.\ each \texttt{HEALPix} cell (that is not excluded by the survey mask) is a member of exactly one subsample. We proceed by computing the angular power spectra omitting one subsample at a time. Calling the angular power spectrum obtained by omitting the $i$th subsample $\left\lbrace C_\ell^{XY}\right\rbrace_i$, we can estimate the covariance matrix as \citep[e.g.][]{Norberg:2008tg}
\begin{equation}
    \widehat{\mathbfss{S}}_{\ell\ell^\prime}^{WXYZ} = \frac{N_\mathrm{sub} - 1}{N_\mathrm{sub}}\,\sum_{i = 1}^{N_\mathrm{sub}}\,\left(\left\lbrace C_\ell^{WX}\right\rbrace_i - \bar C_\ell^{WX}\right)\,\left(\left\lbrace C_{\ell^\prime}^{YZ}\right\rbrace_i - \bar C_{\ell^\prime}^{YZ}\right)\;,
    \label{eq:jacknife_cov}
\end{equation}
where 
\begin{equation}
    \bar C_\ell^{XY} = \frac{1}{N_\mathrm{sub}}\,\sum_{i = 1}^{N_\mathrm{sub}}\left\lbrace C_\ell^{XY}\right\rbrace_i
\end{equation}
is the mean of the angular power spectrum over all subsamples, and the prefactor in equation \eqref{eq:jacknife_cov} comes from the fact that $N_\mathrm{sub} - 2$ pixel groups are the same between each pair of subsamples, thus, one has to correct the covariance matrix estimate for the correlation between each pair of $\left\lbrace C_\ell^{XY}\right\rbrace_i$.

\subsubsection{Sample Covariance of Mock Realisations}
The first external covariance estimator is the most straight forward and most used one. Having generated $N_\mathrm{mock}$ mock realisations of the data as described in \autoref{sec:mocks}, one can simply compute the sample covariance as 
\begin{equation}
    \widehat{\mathbfss{S}}_{\ell\ell^\prime}^{WXYZ} = \frac{1}{N_\mathrm{mock} - 1}\sum_{i = 1}^{N_\mathrm{mock}}\left(\left\lbrace C_\ell^{WX}\right\rbrace_i - \bar C_\ell^{WX}\right)\left(\left\lbrace C_{\ell^\prime}^{YZ}\right\rbrace_i - \bar C_{\ell^\prime}^{YZ}\right)\;.
    \label{eq:sample_cov}
\end{equation}
Here, one has to be aware of the fact that even though \autoref{eq:sample_cov} is an unbiased estimator of the covariance matrix, this is not true for its inverse, the precision matrix $\mathbfss{K}_{\ell\ell^\prime}^{WXYZ}\equiv \left(\mathbfss{S}^{WXYZ}\right)^{-1}_{\ell\ell^\prime}$ which is actually the crucial quantity for inference purposes. An unbiased estimator of the $p\times p$ precision matrix is given by \citep*{Kaufman, Hartlap:2006kj}
\begin{equation}
    \widehat{\mathbfss{K}}^{WXYZ}_{\ell\ell^\prime} = \frac{N_\mathrm{mock} - p - 2}{N_\mathrm{mock} - 1} \left(\widehat{\mathbfss{S}}^{WXYZ}\right)^{-1}_{\ell\ell^\prime}\;.
    \label{eq:hartlap}
\end{equation}

\subsubsection{Covariance from Mock Realisations using the Graphical Lasso}
As we are primarily interested in the precision matrix, we can also apply an estimator designed to directly find sparse precision matrices from realisations of the data. Such an estimator is the graphical lasso \citep*{lasso}. The algorithm works by finding the non-negative definite matrix $\widehat{\mathbfss{K}}^{XYXY}_{\ell\ell^\prime}$ that minimises the log-likelihood of the mock realisations. The strength of the graphical lasso is recovering the graphical structure from correlations in the data. This works better for the inverse correlation matrix $\widehat{\mathbfss{R}}^{XY}_{\ell\ell^\prime}\equiv \widehat{\mathbfss{R}}^{XYXY}_{\ell\ell^\prime}$ than for the precision matrix $\widehat{\mathbfss{K}}^{XYXY}_{\ell\ell^\prime}$. We get the precision matrix as $\widehat{\mathbfss{K}}^{XY}_{\ell\ell^\prime} = \widehat{\mathbfss{R}}^{XY}_{\ell\ell^\prime}/(\sigma_\ell\sigma_{\ell^\prime})$, where $\sigma_\ell\equiv \sqrt{\left\langle \left(\left\lbrace C_\ell^{XY}\right\rbrace_i - \bar C_\ell^{XY}\right)^2\right\rangle}$ is the standard deviation of the angular power spectra estimated from the mocks. As covariance, precision and correlation matrices are usually sparse, there is also a penalty term on off-diagonal terms. The full cost function with the penalty term reads
\begin{equation}
    -\ln\det \widehat{\mathbfss{R}}^{XY} + \sum_{\ell\ell^\prime}\left[\sum_{i=1}^{N_\mathrm{mock}}S_\ell^i\widehat{\mathbfss{R}}^{XY}_{\ell\ell^\prime} S_{\ell^\prime}^i + \lambda \left\vert \widehat{\mathbfss{R}}^{XY}_{\ell\ell^\prime}\right\vert \left(1 - \delta_{\ell \ell^\prime}^\mathrm{K}\right)\right]\;,
    \label{eq:lasso}
\end{equation}
where 
\begin{equation}
    S_\ell^i\equiv \frac{\left\lbrace C_\ell^{XY}\right\rbrace_i - \bar C_\ell^{XY}}{\sigma_\ell}
\end{equation}
are the standardised angular power spectra and $\lambda$ is a hyperparameter that describes the assumed noisiness of the off-diagonal terms. In the limit of $\lambda = 0$, thus assuming the off-diagonal terms of the sample covariance to be noise-free, one can show that equation \eqref{eq:hartlap} minimises equation \eqref{eq:lasso}. We use the graphical lasso implementation of the \verb~scikit-learn~ python package \citep{scikit-learn}, which also includes a cross validation method to automatically choose the value for $\lambda$.

\subsubsection{Comparison of Covariance Matrices}
\label{sec:covcomparison}

\begin{figure*}
    \centering
    \includegraphics[width = 0.7\textwidth]{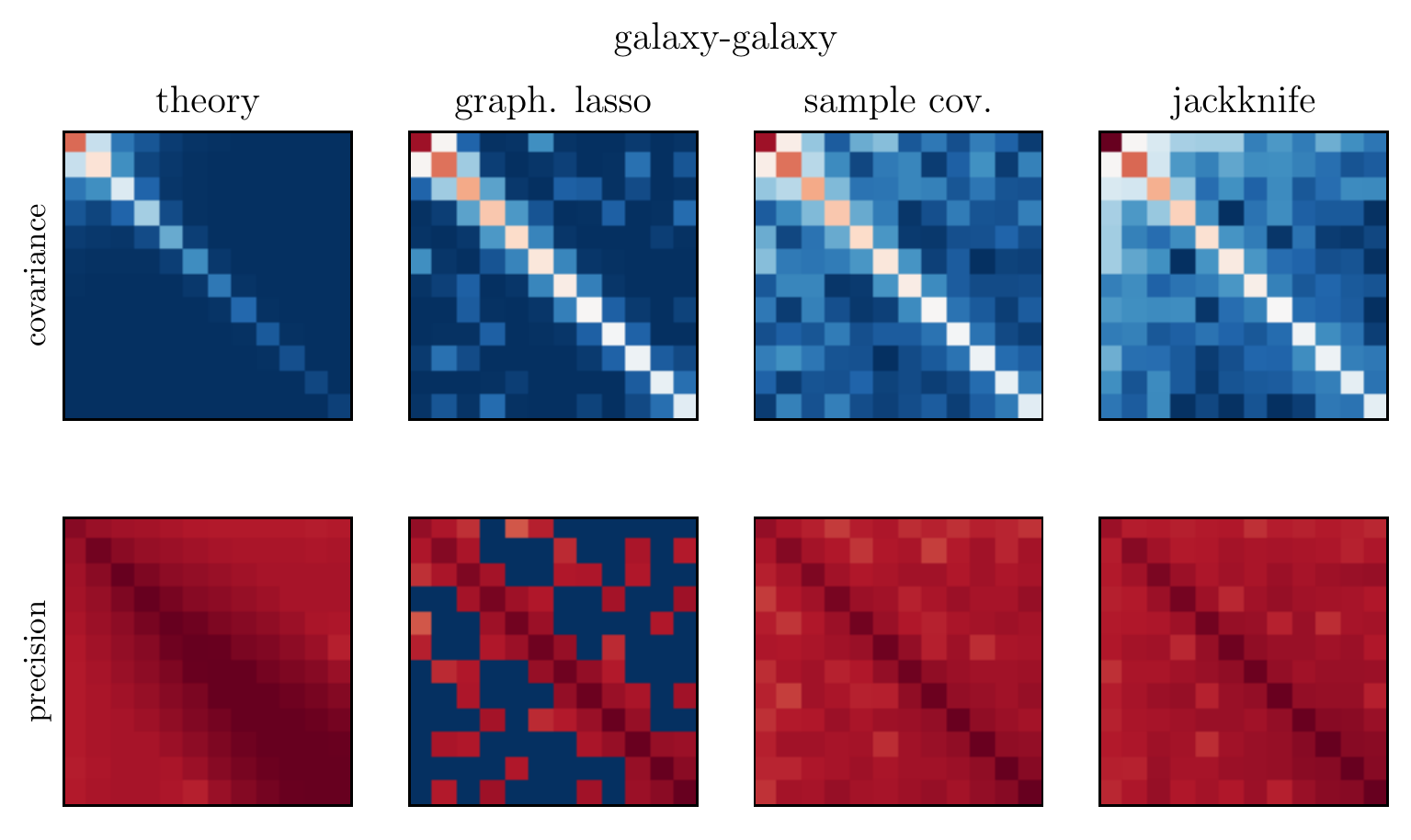}
    \includegraphics[width = 0.7\textwidth]{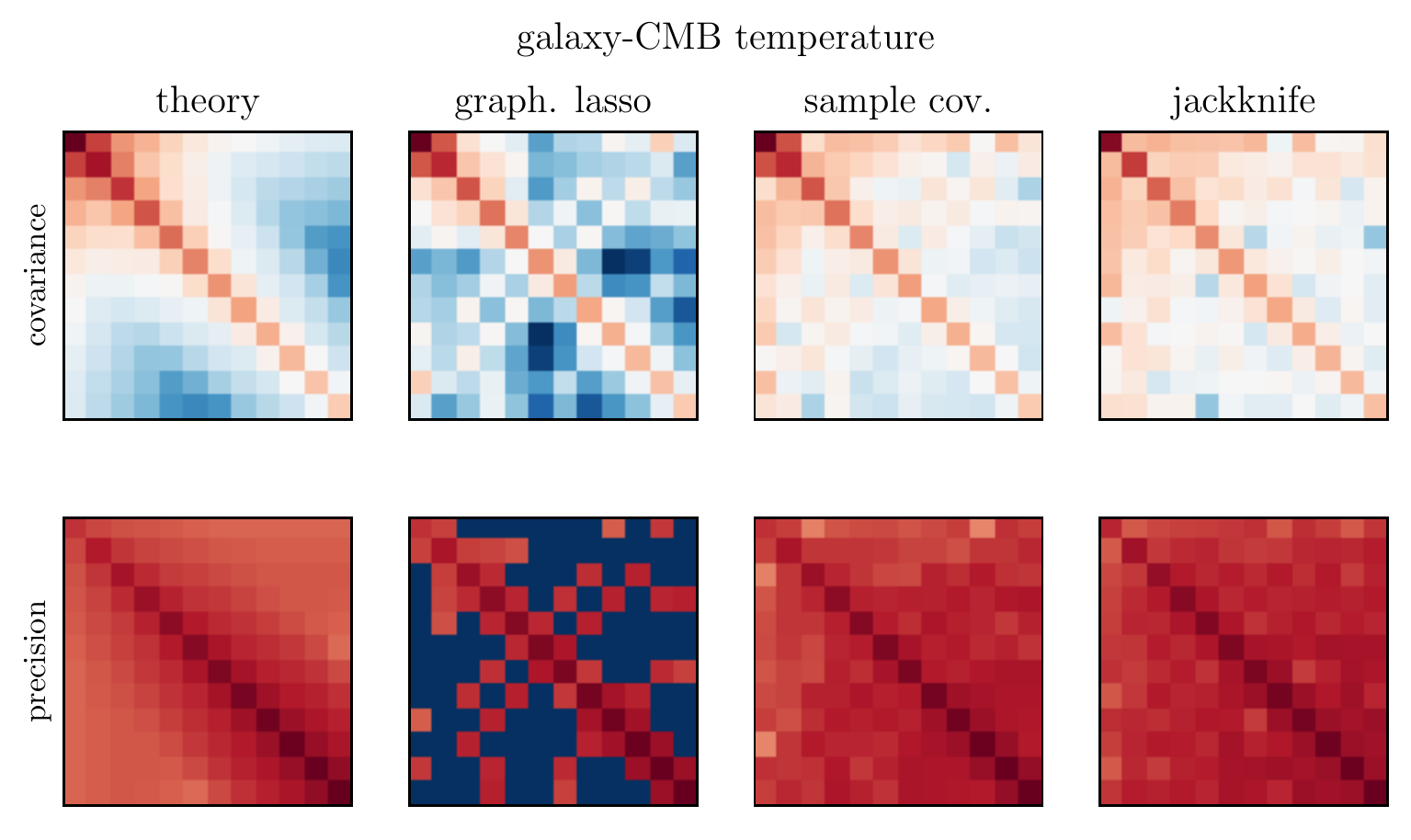}
    \caption{Comparison of the absolute values of the covariance (top line) and precision (bottom line) matrices for $C_\ell^\mathrm{gg}$ and $C_\ell^\mathrm{gT}$ obtained analytically/theoretically, from mock data using the graphical lasso algorithm and by computing their sample covariances, and from jackknife resampling. The variance in the lowest $\ell$-bin is shown in the top-left of each panel, while $\ell$ increases towards the right and bottom, with the bins matching those of the measured power spectrum. The colour scaling is logarithmic.}
    \label{fig:cov_comparison}
\end{figure*}

We plot the covariance and precision matrices obtained with the above-mentioned estimators in \autoref{fig:cov_comparison}. There is reasonable agreement among all of them, though one can spot some significant differences:
\begin{itemize}
    \item The analytic galaxy-galaxy covariance shows smaller values on the diagonal at small scales as those obtained using numerical methods. 
    \item The graphical lasso variances agree well with the sample variance and the jackknife variance. The off-diagonal values are smaller, which is expected as the method is set up to find sparse matrices. For the ISW covariance, the off-diagonal terms are smaller than the analytic prediction though, which hints at a too large value of the hyperparameter $\lambda$. However, increasing $\lambda$ would also increase the suspicious lines of increased covariance perpendicular to the diagonal that are also prominent in the precision matrix.
    \item The sample covariance matrix agrees on the diagonal well with the graphical lasso estimates, whereas the off-diagonal entries look like the analytic covariance matrix with added noise, as expected.
    \item The jackknife resampling method slightly underestimates the galaxy-temperature covariance at large scales, which is expected as the method is inherently blind to cosmic variance. However, for the galaxy-galaxy covariance, jackknife resampling yields larger estimates of the covariance at large scales, which is because all other methods make use of a model whereas the data shows a large-scale power offset compared to our fiducial model that we further discuss in the following section. At smaller scales, the jackknife covariance agrees remarkably well with the sample covariance.
\end{itemize}

As we shall later justify, ignoring the gg power spectrum at large scales, we use the sample covariance of our mocks to attain the main results of this article because it absorbs effects from the survey window, does not rely on any hyperparameters and embodies cosmic variance in the large scale gT power spectrum.

\begin{figure}
    \centering
    \includegraphics[width = \columnwidth]{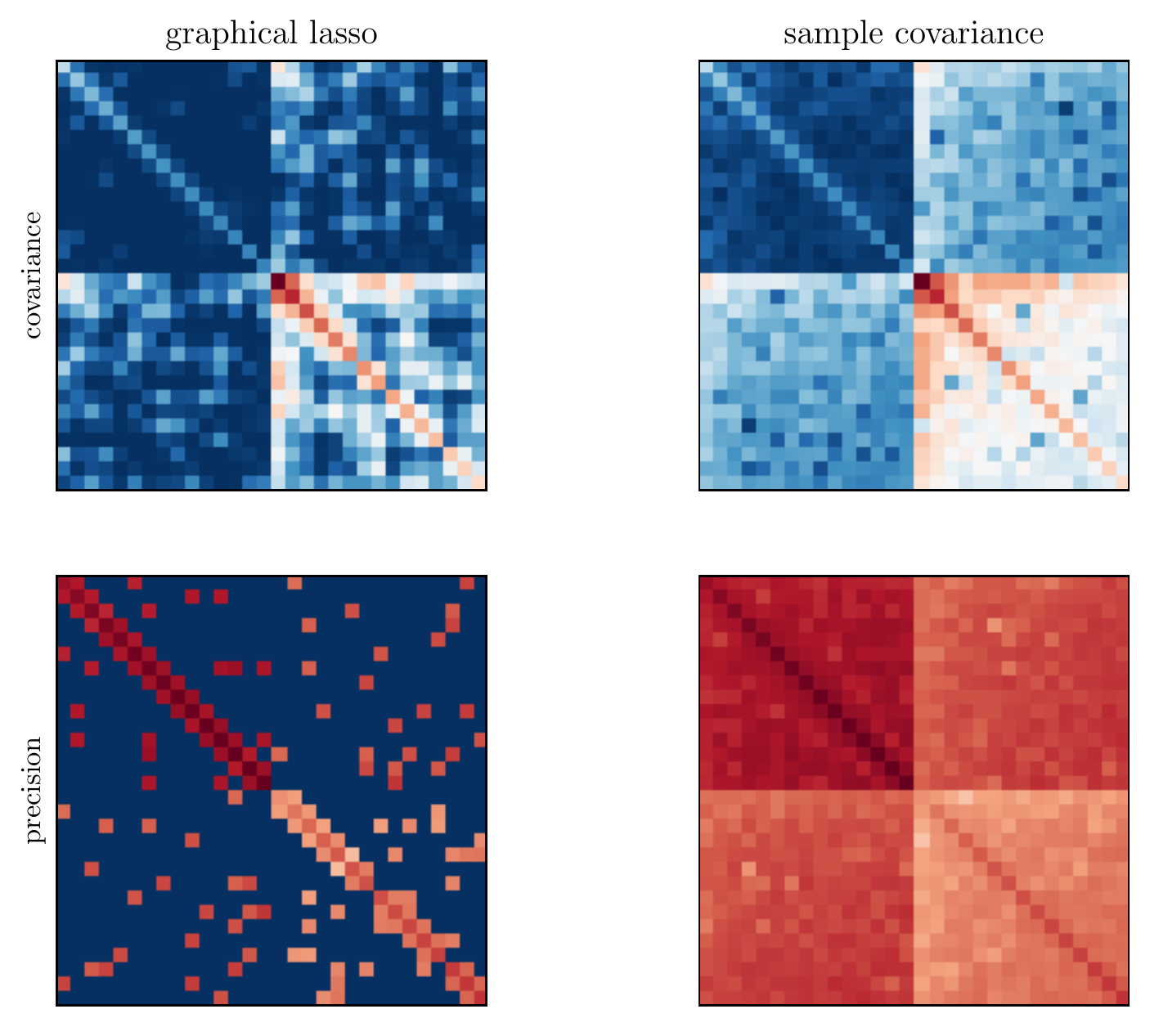}
    \caption{Covariance and precision matrices estimated from data vectors containing both galaxy auto-power spectra and galaxy-temperature power spectra. The top left block is the galaxy-galaxy covariance/precision, the bottom right block the galaxy-temperature submatrix and the top right and bottom left blocks show the galaxy-galaxy-galaxy-temperature covariance/precision.}
    \label{fig:gg_isw_cov}
\end{figure}

When performing a joint analysis of the galaxy-galaxy and galaxy-temperature power spectra, we generally have to take the galaxy-galaxy-galaxy-temperature covariance into account. We have estimated $\widehat{\mathbfss{S}}_{\ell\ell^\prime}^\mathrm{gggT}$ from mock realisations only because we do not have a reliable analytic model for it and jackknife realisations have little advantage here, as cross-correlations are mostly unaffected by observational systematic errors such as foregrounds. Our estimated galaxy-galaxy-galaxy-temperature covariance and precision matrices are visualised in \autoref{fig:gg_isw_cov}. By eye, we cannot identify any particular features in the $\widehat{\mathbfss{S}}_{\ell\ell^\prime}^\mathrm{gggT}$ estimated using the graphical lasso method. In the sample covariance, one can make out a slight increase on the diagonal at large scales, but above the first 5 $\ell$-bins, we do not see any difference between diagonal and off-diagonal terms, raising the suspicion that these matrices are dominated by noise rather than actual correlations. Fortunately, using the full matrix shown in \autoref{fig:gg_isw_cov} provides an equivalent value of $\chi^2$ as when dropping gggT correlations in the $\chi^2$ computation. We henceforth set $\widehat{\mathbfss{S}}_{\ell\ell^\prime}^\mathrm{gggT} = 0$ for all $\ell$ and $\ell^\prime$. 

\subsection{Markov Chain Monte Carlo Sampling}
\label{sec:mcmc}

Even though the angular power spectrum is not normally distributed at large scales \citep[e.g.][]{WMAP:2003pyh,Percival:2006ss}, the distribution of angular power spectra measured from our mock catalogues is approximately Gaussian when binning in relatively wide bins of width $\Delta\ell = 20$ due to the central limit theorem. We, therefore, conjecture the likelihood of the data $\tilde{C}_\ell$ given the model $C_\ell$ as (ignoring the constant normalisation term)
\begin{equation}
    -2\ln\mathcal{P}\left(\left.\tilde{C}_\ell\right\vert C_\ell, \widehat{\mathbfss{K}}_{\ell\ell^\prime}\right)\\ = \sum_{X\in\lbrace \mathrm{gg}, \mathrm{gT}\rbrace} \sum_{\ell \ell^\prime} \Delta C_\ell^{X} \widehat{\mathbfss{K}}_{\ell\ell^\prime}^{XX} \Delta C_{\ell^\prime}^{X}
\end{equation}
with $\Delta C_\ell^\mathrm{X} = \tilde{C}_\ell^{X} - C_\ell^{X}$. Note that we assume here that the shot noise has already been subtracted from $\tilde{C}_
\ell$ (cf. \autoref{sec:estimators}).

Before using this likelihood in any Markov chain Monte-Carlo (MCMC) sampling, we evaluate it for our fiducial model with and without putting the Limber approximation into service. A difference occurs only at the third significant digit which justifies our reliance on the Limber approximation to avoid our MCMC sampling being considerably more expensive.

We sample the posterior distribution of the parameters of interest using \textit{Ensemble Slice Sampling} \citep{karamanis2020ensemble} implemented in the \verb~zeus~ code \citep*{karamanis2021zeus}. For one parameter, given a starting point $x_0$ and calling the probability density function to be sampled $f(x_0)$, \textit{Slice Sampling} works by iterating over the following steps \citep{NealSliceSampling}:
\begin{enumerate}
    \item Draw a uniformly distributed height $y_i$ from the interval $[0, f(x_i)]$;
    \item Define the slice $S=\lbrace x:y_i < f(x)\rbrace$;
    \item Uniformly draw a new point $x_{i + 1}$ from $S$.
\end{enumerate}
The advantages of this sampler compared to many other MCMC samplers are that one does not have to define any proposal distribution for efficient application (it is a so-called black box) and that its acceptance rate is 1. On the downside, the Slice Sampler has to evaluate $f(x)$ multiple times per step to numerically approximate the slice interval $S$.

For more than one parameter, each slice $S$ has as many dimensions as parameters, thus, one has to define a direction along which the next point ${\bm x}_{i + 1}$ is chosen. \verb~zeus~ runs an ensemble of Slice Samplers in parallel, and, by default, the new point ${\bm x}_{i + 1}^\mathhangeulga$ of the $\mathhangeulga$th walker is chosen along the vector
\begin{equation}
    {\bm\eta}_\mathhangeulga = \mu \left({\bm x}_{i}^\mathhangeulna - {\bm x}_{i}^\mathhangeulda\right),
    \label{eq:ESS_vector}
\end{equation}
where ${\bm x}_{i}^\mathhangeulna$ and ${\bm x}_{i}^\mathhangeulda$ are the current position of two walkers other than $\mathhangeulga$ drawn uniformly and without replacement, and $\mu$ is a length scale that, as the sampling progresses, is tuned to reduce the number of $f(\mathrm{x})$ evaluations needed to find the slice interval. As the distribution of walkers, after a burn-in period, resembles the target distribution, Eq.~\eqref{eq:ESS_vector} naturally prefers directions of correlated parameters \citep{karamanis2020ensemble}.

We employ \verb~ChainConsumer~ \citep*{samuel_hinton_2020_4280904} to analyse our chains.

\section{Results}
\label{sec:results}

\subsection{The Galaxy-Galaxy Auto-Power spectrum}
\label{sec:cl_result}

\begin{figure}
	\includegraphics[width=\columnwidth]{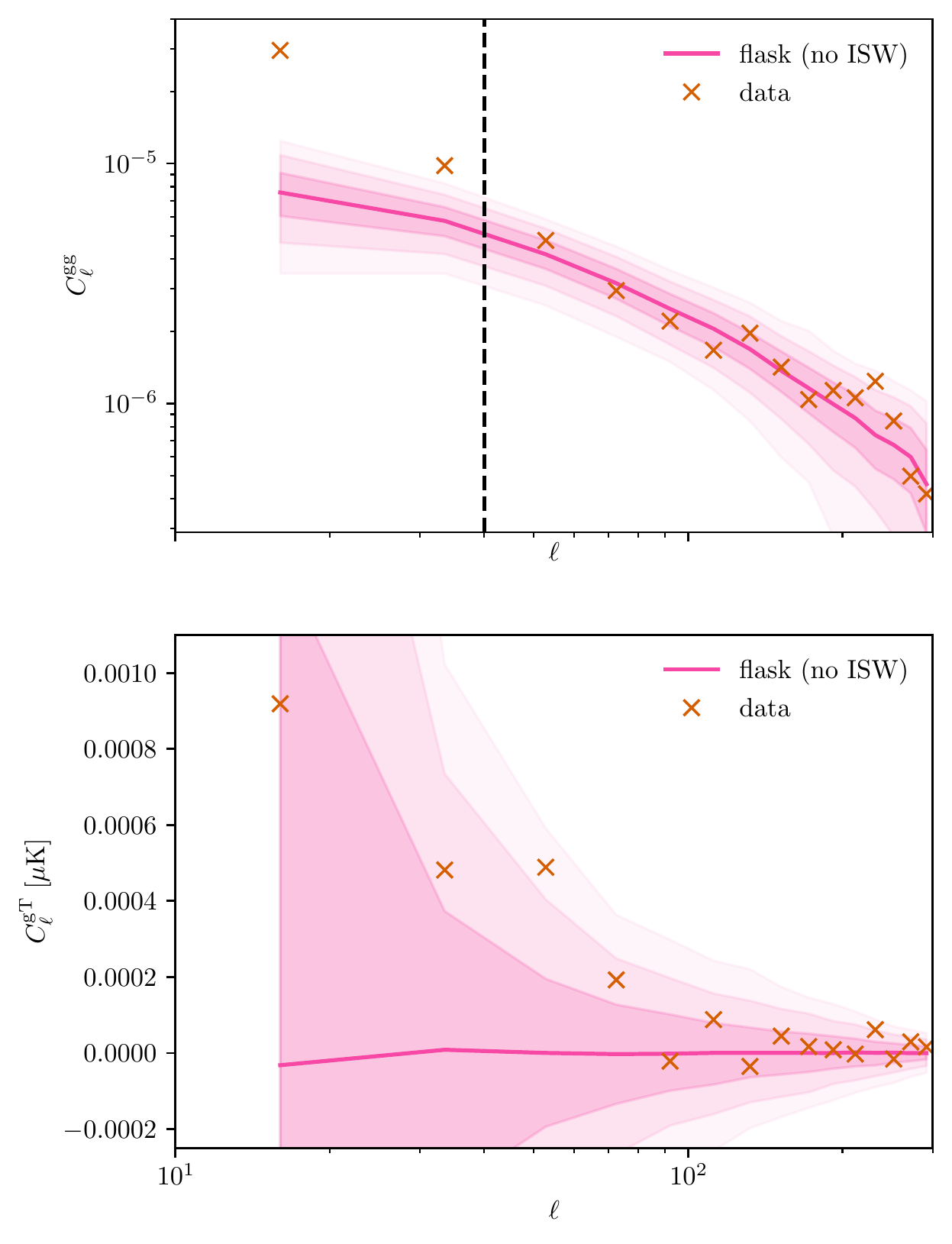}
    \caption{The angular auto-power spectra $\tilde{C}_\ell^\mathrm{gg}$ measured from the RACS island catalogue (crosses, top) and the RACS-Planck cross-power spectrum $\tilde{C}_\ell^\mathrm{gT}$ (bottom). The magenta line shows the median $C_\ell$ of the flask realisations and the shaded regions show the 68-, 95- and 99.75-percentile regions. In the top plot, we mark $\ell = 40$ as the upper bound of the distrusted multipole range that we do not include in our analyses of $C_\ell^\mathrm{gg}$.}
    \label{fig:cl_RACS}
\end{figure}

In \autoref{fig:cl_RACS}, we show the measured angular galaxy auto-power spectrum $\tilde{C}_\ell^\mathrm{gg}$ in $\ell$-bins with width $\Delta\ell = 20$ for a flux limit of 4 mJy. We also plot the fiducial power spectrum that we use to set up \verb~flask~ along with percentile regions estimated from 3000 \verb~flask~ realisations. We see a good agreement of the fiducial model with the data at $\ell>40$. At larger scales, however, we see more power than expected. We suspect that this offset is due to hitherto unidentified systematic effects and discuss this further below and in Appendix \ref{sec:lsssyst}. In  \citet{2021MNRAS.502..876A}, the angular clustering data from LOFAR on scales larger than the size of a pointing was removed  due to systematic effects. 

\begin{figure}
	\includegraphics[width=\columnwidth]{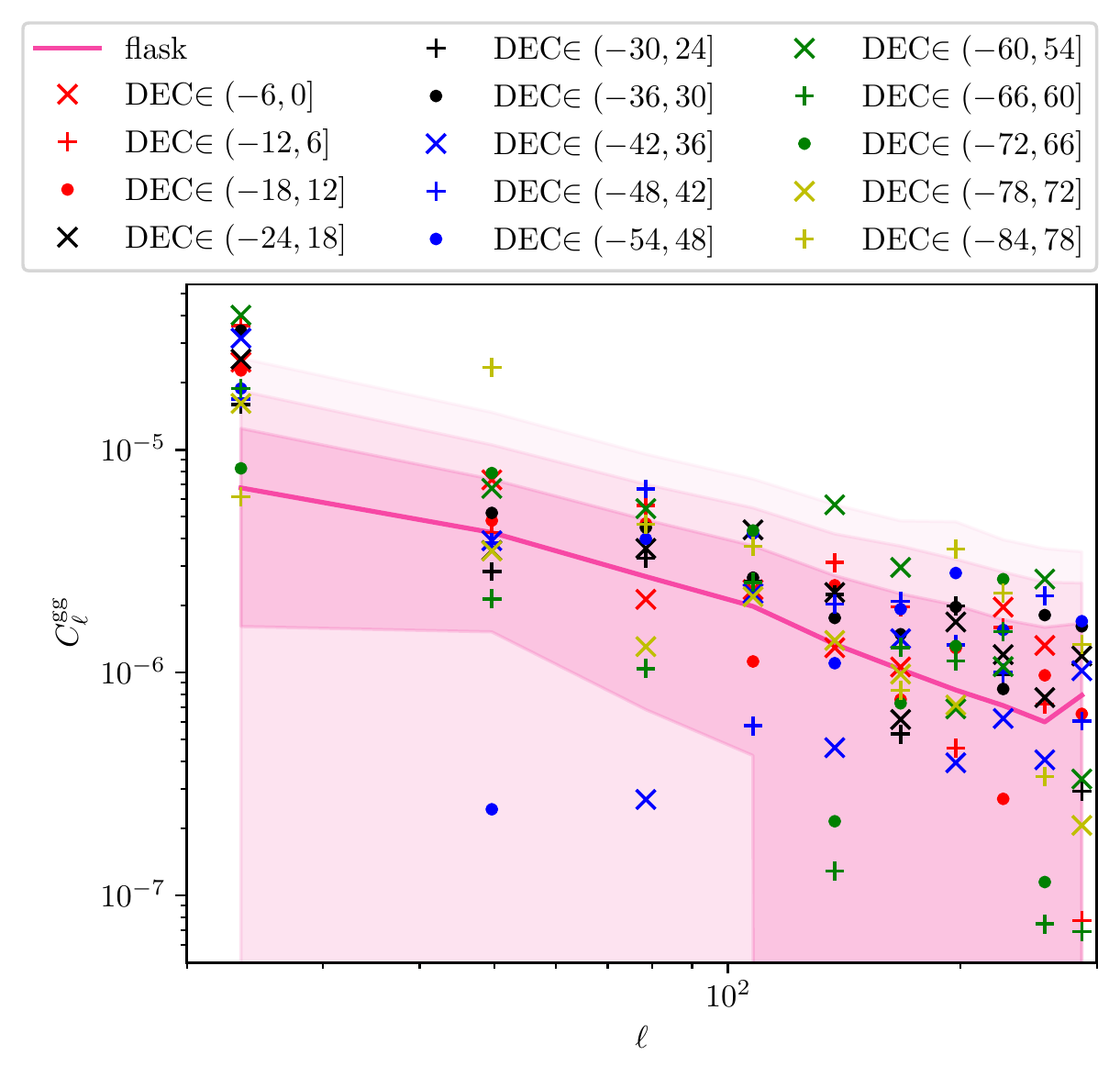}
	\includegraphics[width=\columnwidth]{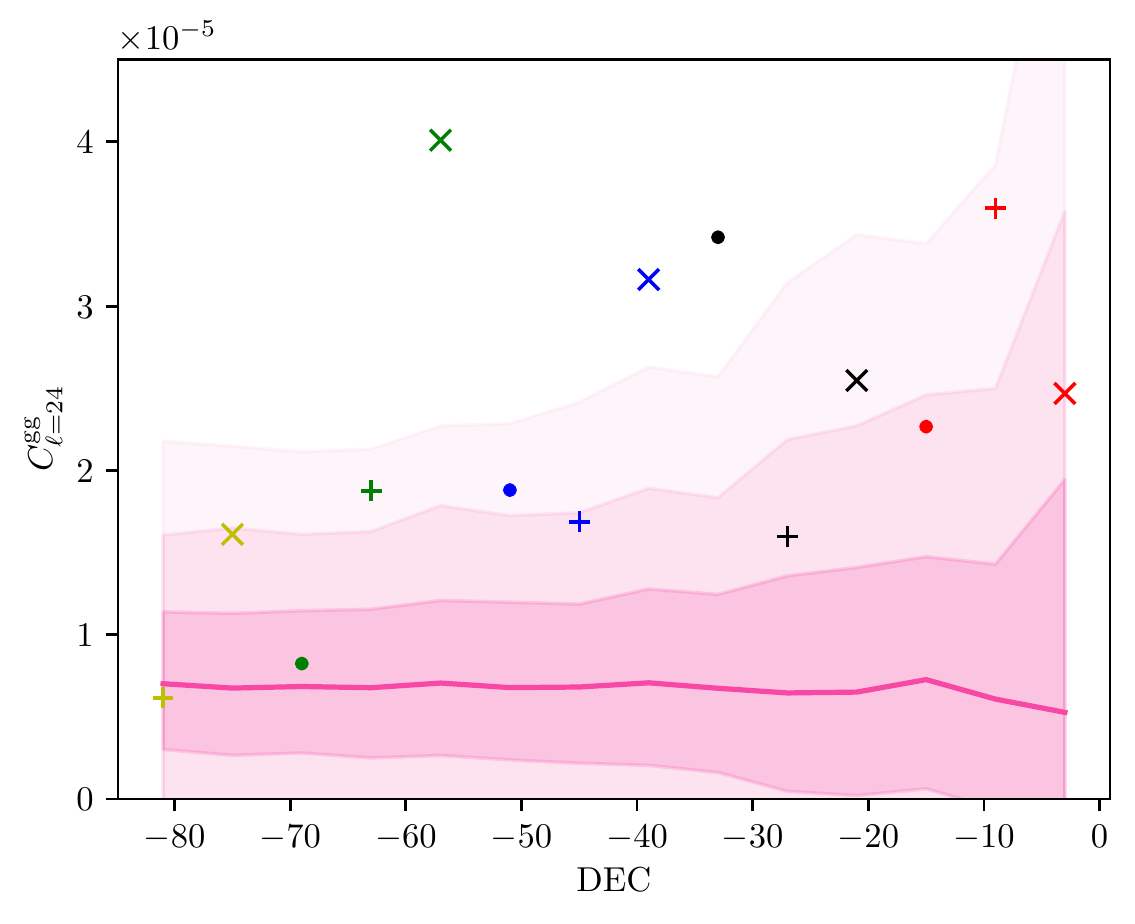}
    \caption{The top panel shows the $\tilde{C}_{\ell}^\mathrm{gg}$ measured in declination (DEC) bands with widths of 6 degrees. The bottom panel shows the $\tilde{C}_{\ell}^\mathrm{gg}$ at $\ell=24$ as a function of DEC. The marker colours and shapes coincide in both plots. The magenta line and shaded regions show the median and 68-, 95- and 99.75-percentile regions of the flask realisations, as in \autoref{fig:cl_RACS}.}
    \label{fig:Cl_DECcut}
\end{figure}

The assumption that this large-scale power excess is due to systematic effects is further supported by the fact that when we measure the galaxy-galaxy auto-power spectrum in stripes of constant declination  with a width of 6 degrees, we see less power at the largest scales in stripes  that are closer to the South Pole (cf. \autoref{fig:Cl_DECcut}). Interestingly, our mock catalogues suggest that the error on $\tilde C_{\ell = 24}^\mathrm{gg}$ increases towards the equator as well, regardless of the increased area subtended by the declination strip. Since the only direction dependent information that enters the generation of the mock catalogues is the radio data weighting function $w(\btheta)$, we suspect this unexpected behaviour to be due to an increased number of pixels where $w(\btheta)$ is low as we go further north. This shall be studied in more detail in future work in preparation for the EMU survey.

Despite this behaviour that is correlated with declination, we cannot simply ignore data on the fact that they do not match our expectations. We will therefore perform a first bias measurement both with and without considering large-scale (i.e.\ $\ell \leq 40$) galaxy clustering data. The measured bias parameters for both $n(z)$ models and bias models \ref{item:expcutbias}-\ref{item:constbias} are tabulated in \autoref{tab:b_evo_results}. These have been obtained by simple numerical optimisation methods and thus are reported without errors which we deliver later (cf. Tables \ref{tab:b_AISW_results} and \ref{tab:bias_models_AISW_results}) after running MCMC jointly on the gg and gT power spectra. The aim here is to check how well our modelling assumptions can describe the data.

When we include multipoles at $\ell \leq 40$, the galaxy bias (for non-constant bias parameterisations) surprisingly decreases with redshift. Furthermore, the minimum $\chi^2$ is from three to more than twelve times larger than the number of degrees of freedom, suggesting that our model is insufficient at large scales. We, therefore, make use of the galaxy auto-power spectrum at $\ell > 40$ only (unless otherwise stated) and leave it to be reanalysed in the future when either an extended model or a better understanding of systematic effects is at hand.

\begin{figure}
    \centering
    \includegraphics[width = \columnwidth]{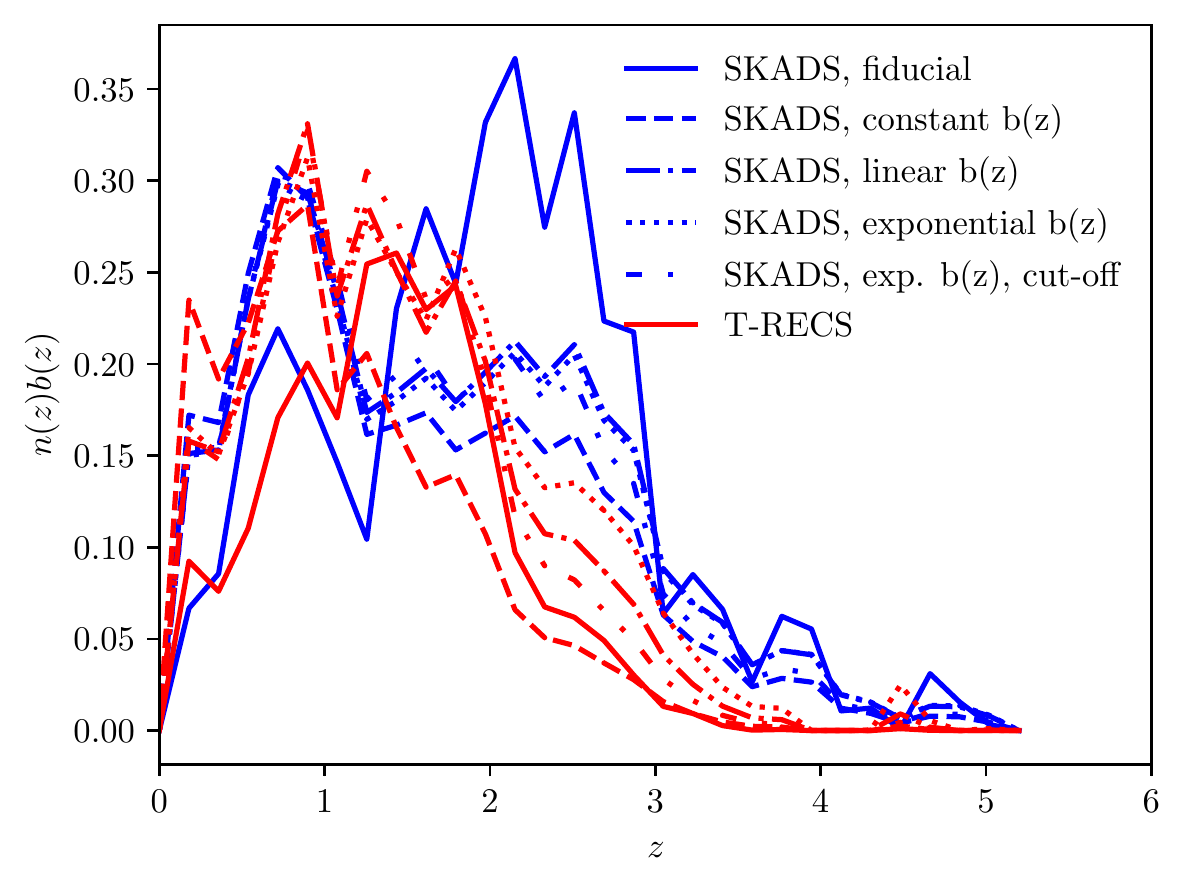}
    \caption{The product of the SKADS/T-RECS $n(z)$ distribution and the best-fitting galaxy biases $b(z)$. The solid blue line shows $n(z)b(z)$ for when we use the fiducial bias parameters used in \citet{Raccanelli:2014kga,Bernal:2018myq} and \citet{Asorey:2021hgc}. The solid red line shows the estimate of $n(z)b(z)$ from T-RECS (cf. \autoref{fig:bias_trecs2})}
    \label{fig:nbz}
\end{figure}

Omitting large-scale multipoles at $\ell \leq 40$, we find almost equal values of $\chi^2/\mathrm{dof}$ for all bias parameterisations and both $n(z)$ models, with the exception of using the T-RECS $n(z)$ with a constant bias. This model stands out in \autoref{fig:nbz} as the one where $n(z)b(z)$ drops quite sharply above $z>1$, whereas other T-RECS models have a wider peak region that extends up to $z\sim 2$ and the $n(z)b(z)$ of best-fitting SKADS models have a peak similar to the constant-bias T-RECS model but have a plateau between $1.4 \lesssim z \lesssim 2.6$ such that, in this redshift range, the average $n(z)b(z)$ is the same as for the T-RECS models with bias evolution. Yet, even in the constant-bias T-RECS case, $\chi^2/\mathrm{dof}$ is much lower than in any full $\ell$-range case. In all other cases, $\chi^2/\mathrm{dof}$ is only marginally greater than unity, implying that all of these models describe the data well. Instead of trying to choose one particular model, we shall use the scatter of the results obtained with these different models to estimate the systematic uncertainty.

In any case, neither the best-fitting parameters nor the minimum $\chi^2$ show much difference between the pure exponential bias parameterisation \ref{item:expbias} and its variant \ref{item:expcutbias} with a constant bias above $z > 1.5$. Considering this result and the fact that we regard the redshift cut as arbitrary, we do not pursue model \ref{item:expcutbias} any further.

\begin{table}
    \centering
    \caption{Maximum posterior values of the galaxy bias $b_0$ at redshift $z = 0$ and the bias's redshift evolution from minimising the $C_\ell^\mathrm{gg}$ $\chi^2$. The effective redshift $z_\mathrm{eff}$ has been obtained by integrating over $z b(z) n(z)$ for the best-fitting bias parameters. The bias $b(z)$ is included in the $z_\mathrm{eff}$ integral as it is degenerate with $n(z)$, and, therefore, our $b(z)$ measurement is, to some degree, also effectively accounts for potential deviations from the fiducial $n(z)$ distribution.}
    \label{tab:b_evo_results}
    \begin{tabular}{rcccccc}
		\hline
		Model & $b_0$ & $\frac{\mathrm{d}b}{\mathrm{d}z}$ or  & $z_\mathrm{eff}$ 
		& $b(z_\mathrm{eff})$ & $\chi^2_\mathrm{min}$ & $\chi^2_\mathrm{min}/\mathrm{dof}$\\ & &  $\frac{\mathrm{d}\ln b}{\mathrm{d}z}$ & & & \\ 
		\hline \hline
		\multicolumn{5}{c}{All $\ell$}\\
		\multicolumn{5}{l}{SKADS}\\
		const. $b(z)$ & $3.63$ & $-$ & $1.56$ & $3.63$ &  $172.8$ & $12.34$ \\
		linear $b(z)$ & $4.25$ & $-0.85$ & $1.28$ & $3.16$ &  $131.6$ & $10.12$ \\
		exp. $b(z)$ & $7.80$ & $-1.79$ & $0.65$ & $2.44$ & $45.27$ & $3.482$ \\
		- w/cut-off & $7.79$ & $-1.79$ & $0.88$ &1.61 & $45.10$ & $3.469$ \\
        \hline
		\multicolumn{5}{l}{T-RECS}\\
		const. $b(z)$ & $2.66$ & $-$ & $1.13$ & $2.66$ & $93.85$ & $6.703$ \\
		linear $b(z)$ & $2.95$ & $-0.59$ & $1.00$ & $2.36$ & $72.05$ & $5.542$ \\
		exp. $b(z)$ & $3.96$ & $-1.11$ & $0.72$ & $1.78$ & $45.83$ & $3.525$ \\
		- w/cut-off & $3.96$ & $-1.11$ & $0.65$ & $1.92$ & $45.72$ & $3.517$ \\
		\hline \hline
		\multicolumn{5}{c}{Only $\ell > 40$}\\
		\multicolumn{5}{l}{SKADS}\\
		const. $b(z)$ & $3.24$ & $-$ & $1.56$ & $3.24$ & $12.70$ & $1.058$ \\
		linear $b(z)$ & $2.74$ & $0.59$ & $1.72$ & $3.75$ & $12.11$ & $1.101$ \\
		exp. $b(z)$ & $2.83$ & $0.15$ & $1.72$ & $3.66$ & $12.11$ & $1.101$ \\
		- w/cut-off & $2.71$ & $0.23$ & $1.64$ & $3.62$ & $12.09$ &  $1.099$ \\
        \hline
		\multicolumn{5}{l}{T-RECS}\\
		const. $b(z)$ & $2.41$ & $-$ & $1.13$ & $2.41$ & $17.19$ & $1.433$ \\
		linear $b(z)$ & $1.33$ & $1.62$ & $1.40$ & $3.60$ & $11.78$ & $1.071$ \\
		exp. $b(z)$ & $1.52$ & $0.61$ & $1.51$ & $3.82$ & $11.71$ & $1.065$ \\
		- w/cut-off & $1.43$ & $0.73$ & $1.34$ & $3.80$ & $11.67$ & $1.061$ \\
		\hline
    \end{tabular}
\end{table}

\subsection{The Galaxy-Temperature Cross-Power spectrum}
\label{sec:gTspectrum}

As the ground-based radio observations of galaxies are subject to very different systematic effects as CMB observations from space, we do not expect significant systematic contributions to the measured cross-power spectrum $\tilde{C}_\ell^\mathrm{gT}$ shown in \autoref{fig:cl_RACS}. We mark again the median and confidence regions estimated from 3000 \verb~flask~ realisations, however, this time, we initialise each simulation to have no intrinsic correlation between the galaxy and CMB map.

We expect most of the ISW signal at large scales. Thus, even though we ignore the first two multipole bins in $C_\ell^\mathrm{gg}$, these are crucial in the $C_\ell^\mathrm{gT}$ analysis. As we show in Appendix \ref{sec:isw_without_large_scales}, it is actually conservative to include large-scale $C_\ell^\mathrm{gT}$ multipoles in the ISW analysis. Assuming that the observed $\tilde{a}_{\ell m}^\mathrm{g} = a_{\ell m}^\mathrm{g} + f_{\ell m}$ is the sum of the true cosmological $a_{\ell m}^\mathrm{g}$ and some unknown systematic $f_{\ell m}$, we have  the observed power spectra
\begin{align}
    \langle \tilde{C}_\ell^\mathrm{gg}\rangle &= \langle \tilde{a}_{\ell m}^{\mathrm{g}}\,\tilde{a}_{\ell m}^{*\mathrm{g}}\rangle = C_\ell^\mathrm{gg} + \left(\langle f_{\ell m}\,a_{\ell m}^{*\mathrm{g}}\rangle + \text{c.c.}\right) + \langle f_{\ell m}\,f_{\ell m}^*\rangle\text{, and}\nonumber\\
    \langle\tilde{C}_\ell^\mathrm{gT}\rangle &= \langle \tilde{a}_{\ell m}^{\mathrm{g}}\,\tilde{a}_{\ell m}^{*\mathrm{T}}\rangle = C_\ell^\mathrm{gT} + \langle f_{\ell m}\,a_{\ell m}^{*\mathrm{T}}\rangle.
\end{align}
If $f_{\ell m}$ is an observational systematic, e.g.\ a terrestrial or Galactic foreground, than it is uncorrelated with the true cosmological signal, i.e.\ $\langle f_{\ell m}\,a_{\ell m}^{*\mathrm{g}}\rangle = \langle f_{\ell m}\,a_{\ell m}^{*\mathrm{T}}\rangle = 0$. Hence, $\langle\tilde{C}_\ell^\mathrm{gT}\rangle = C_\ell^\mathrm{gT}$ is unaffected by the systematic, whereas $\tilde{C}_\ell^\mathrm{gg}$ is biased by the auto-power spectrum of $f_{\ell m}$. On the other hand, if the observed excess is due to a theoretical systematic, i.e.\ it is not predicted well by our modelling of the density field, we will see unexpected behaviour in the gT cross-power spectrum as well. In the latter case, we will see values of $\chi^2$ that exceed the number of degrees of freedom by far. We therefore proceed including the full available multipole range in the gT analysis and will present a simple $\chi^2$ test later to justify this.

Our first step in analysing the significance of the ISW signal in the gT cross-power spectrum is to compare the values of 
\begin{equation}
    \chi^2 = \sum_{\ell, \ell^\prime} \left(\tilde{C}_{\ell}^ \mathrm{gT}-C_{\ell}^ \mathrm{gT}\right)\mathbfss{K}_{\ell\ell^\prime}^\mathrm{gTgT}\left(\tilde{C}_{\ell^\prime}^ \mathrm{gT}-C_{\ell^\prime}^ \mathrm{gT}\right),
\end{equation}
for the two hypotheses of existence and non-existence of gT cross-correlations due to the ISW effect. In the former case $C_{\ell}^ \mathrm{gT}$ is as defined in \autoref{eq:cl_isw}, while in the latter, we just have $C_{\ell}^ \mathrm{gT} = 0$. Using the sample covariance matrix of 3000 mock catalogues, we obtain $\chi^2 = 17.7$ for the null hypothesis ($C_{\ell}^ \mathrm{gT} = 0$) and $\chi^2 = 10.9$ for the $C_\ell^\mathrm{gT}$-model given in \autoref{eq:cl_isw}. If we use instead a precision matrix estimated from the same set of mock catalogues using the graphical lasso method, we find $\chi^2 = 17.8$ for the null hypothesis and $\chi^2 = 11.0$ for ISW hypothesis. So in both cases, adopting an ISW model reduces $\chi^2$ by 6.8. Using the theoretical precision matrix, we obtain $\chi^2 = 10.7$ for the null hypothesis and $\chi^2 = 7.4$ for the ISW hypothesis, underestimating the mode-coupling contribution of the survey mask, and, hence, the significance of the ISW detection. On the contrary, ignoring cosmic variance, the jackknife resampling increases the significance with $\chi^2 = 21.3$ and $\chi^2 = 12.2$ for the null and ISW hypotheses, respectively.

We can further describe the significance of this finding in terms of the signal-to-noise ratio \citep{DES:2015rdu}
\begin{equation}
    \frac{S}{N} = \frac{\sum_{\ell, \ell^\prime}\tilde{C}_{\ell}^ \mathrm{gT}\mathbfss{K}_{\ell\ell^\prime}C_{\ell^\prime}^ \mathrm{gT}}{\sqrt{\sum_{\ell, \ell^\prime}C_{\ell}^ \mathrm{gT}\mathbfss{K}_{\ell\ell^\prime}C_{\ell^\prime}^ \mathrm{gT}}}.
    \label{eq:sn}
\end{equation}
We evaluate Eq. \eqref{eq:sn} again using both covariance matrices estimated from simulations and an $\ell$-binning with $\Delta\ell = 20$ which yields
\begin{equation}
    \boxed{\frac{S}{N} = 2.8}.
\end{equation}
Alternatively, we attain $S/N = 1.9$ with the analytic and $S/N = 3.2$ with the jackknife covariance matrices.

\subsection{Parameter Constraints}

In the previous subsection, we have detected a positive cross-correlation between the galaxy and temperature maps at 2.8 $\sigma$ compared to the null hypothesis of no correlation. However, in the gT cross-power spectrum, the amplitude of the ISW signal $A_\mathrm{ISW}$ is degenerate with the galaxy bias $b(z)$, as well as the redshift distribution of radio continuum sources per steradian $n(z)$. In this section, we reevaluate the significance of our ISW detection taking our ignorance on $b(z)$ and $n(z)$ into account.

\begin{figure}
    \centering
    \includegraphics[width=\columnwidth]{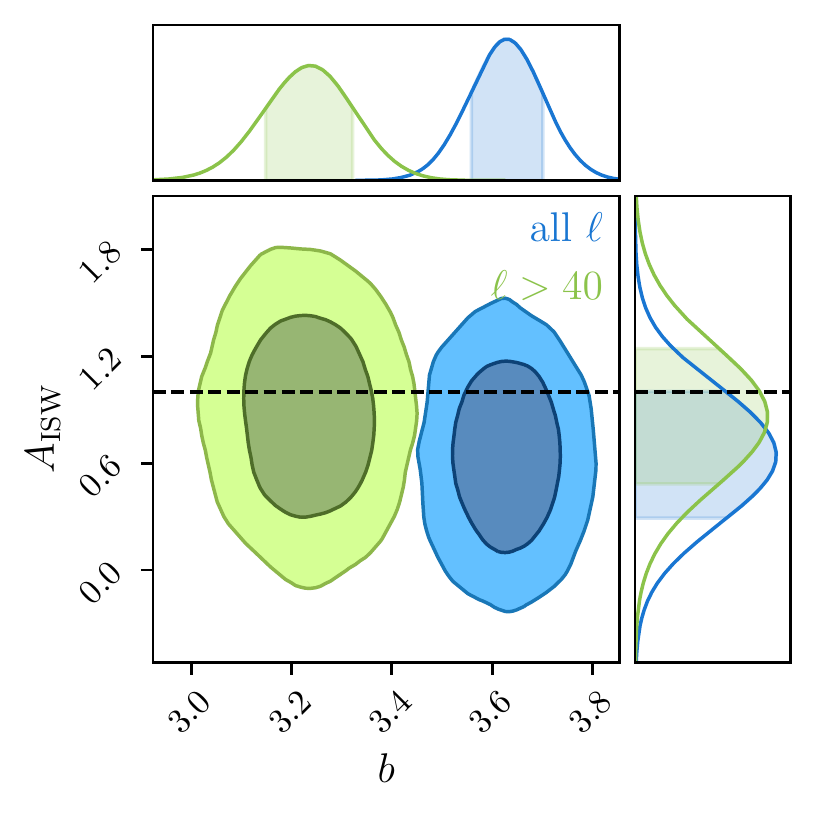}
    \caption{Posterior contours of the galaxy bias $b$ and $A_\mathrm{ISW}$ from jointly analysing $\tilde{C}_\ell^\mathrm{gg}$ and $\tilde{C}_\ell^\mathrm{gT}$ assuming a SKADS $n(z)$. We use the full available $\ell$-range in the gT spectrum, but we omit $\ell \leq 40$ for the green contours. The dashed lines indicates $A_\mathrm{ISW} = 1$. The dark (light) shaded contours contain 68 (95) per cent of the MCMC chain elements. The shaded regions in the histograms correspond to the 68 per cent credible interval.}
    \label{fig:RACS_b_AISW_cornerplot}
\end{figure}

As the gg auto-power spectrum depends only on $b^2(z)n^2(z)$, we can use it to anchor $b(z)$ and, thus, lift the $b(z)$-$A_\mathrm{ISW}$ degeneracy. At the outset, we fix $n(z)$ to the one predicted by SKADS \citep{skads}. We ensemble slice sample a constant bias parameter $b$ and the ISW signal amplitude $A_\mathrm{ISW}$ first using the full measured gg auto-power spectrum, and then repeat the same analysis restricting the gg auto-power spectrum to $\ell > 40$ only, while still taking the full gT cross-power spectrum. The resulting $b$-$A_\mathrm{ISW}$ posterior contours are plotted in \autoref{fig:RACS_b_AISW_cornerplot}. To fit the excess power at low multipoles with our two-parameter model, the galaxy bias $b$ is required to be significantly larger than for the case where we ignore galaxy auto-correlations at $\ell \leq 40$. As we perform both analyses on the same multipole range of the gT cross-power spectrum whose amplitude is given by the product $bA_\mathrm{ISW}$, using the full available range of scales favours smaller values of $A_\mathrm{ISW}$. We are reassured by the fact that the marginalised posteriors on $A_\mathrm{ISW}$ are mostly consistent with each other. The significance of our ISW detection is thus largely unaffected by the large-scale power excess.

The best-fitting values are given in \autoref{tab:b_AISW_results} along with $\chi^2_\mathrm{min}$, the minimum value of $\chi^2$. When omitting large scales in the gg auto-power spectrum, we obtain a reduced $\chi^2$ of 0.97, indicating that our modelling works well to describe the data at these scales. When we include multipoles at $\ell \leq 40$, the reduced $\chi^2$ increases by more than six times the previous value, suggesting that our model is insufficient at large scales. We, therefore, ignore the galaxy auto-power spectrum at $\ell \leq 40$ in the following parts of this article and leave it to be reanalysed in the future when either an extended model or a better understanding of systematic effects is readily available.

\begin{table}
    \centering
    \caption{Maximum posterior values of the galaxy bias $b$ and $A_\mathrm{ISW}$ from jointly analysing $\tilde{C}_\ell^\mathrm{gg}$ and $\tilde{C}_\ell^\mathrm{gT}$ assuming a SKADS $n(z)$. We use the full available $\ell$ range in the gT spectrum, but we omit $\ell \leq 40$ when analysing the gg spectrum for the bottom line.}
    \label{tab:b_AISW_results}
    \begin{tabular}{cccccc}
        \hline
		gg $\ell$-range & $b$ & $A_\mathrm{ISW}$ & $\chi^2_\mathrm{min}$ & dof & $\chi^2_\mathrm{min}/\mathrm{dof}$\\ 
		\hline
		all $\ell$ & $3.613^{+0.087}_{-0.050}$ & $0.68^{+0.32}_{-0.36}$ & 187 & 30 & 6.2\\ 
		$\ell > 40$ & $3.248^{+0.068}_{-0.094}$ & $0.82^{+0.39}_{-0.33}$ & 27.1 & 28 & 0.97\\
		\hline
    \end{tabular}
\end{table}

\begin{figure}
    \centering
    \includegraphics[width = \columnwidth]{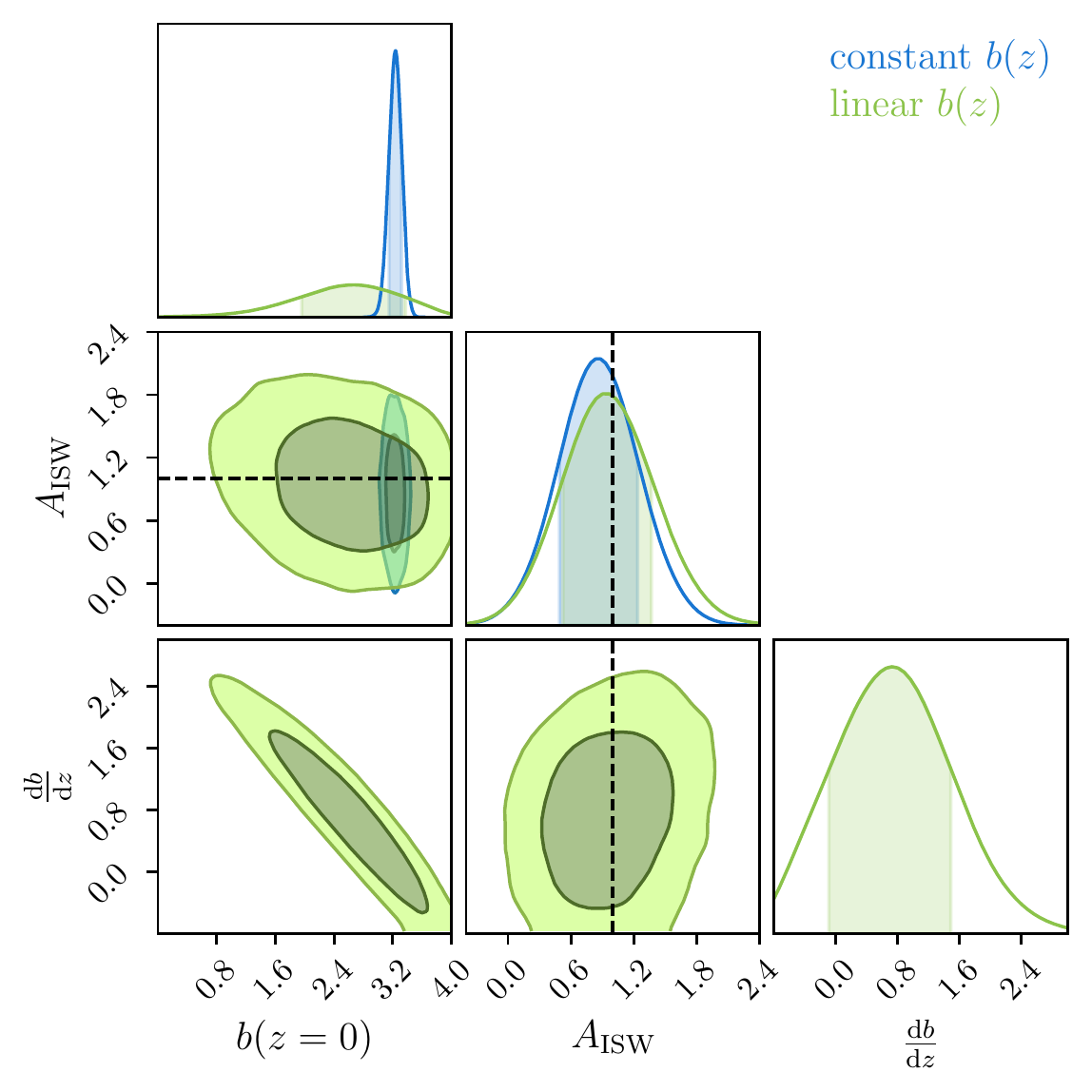}
    \includegraphics[width = \columnwidth]{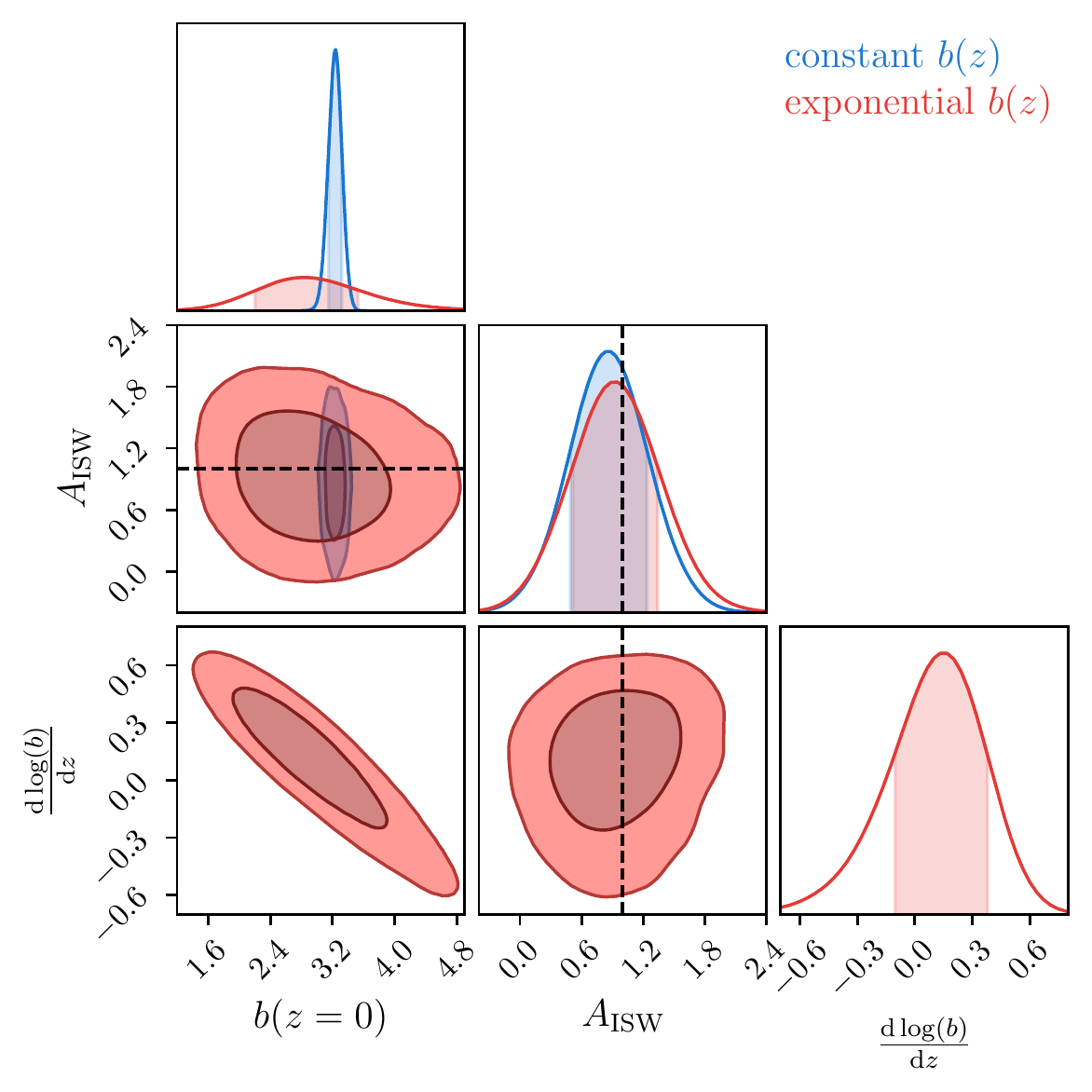}
    \caption{Posterior contours of the galaxy bias $b(z=0)$ at redshift $z=0$, its evolution and $A_\mathrm{ISW}$. In the top plot, we compare the contours for constant and linear bias parameterisations. The bottom panel shows the contours obtained using an exponential bias parameterisation alongside the constant bias contours. All contours shown in this figure have been made without the contributions of multipoles $\ell \leq 40$ to the galaxy-galaxy auto-power spectrum $\tilde{C}_\ell^\mathrm{gg}$. The shading signifies the same as in \autoref{fig:RACS_b_AISW_cornerplot}.}
    \label{fig:bias_contours}
\end{figure}

Our next step is to allow the bias to evolve with redshift. To avoid unphysical results, we adopt an additional prior on combinations of $b(z=0)$ and $\mathrm{d} b/\mathrm{d} z$ in the linear bias case, that is that $b(z) < 0$ is excluded at all redshifts $z$ probed by the survey. This condition is always fulfilled by the exponential bias parameterisation as long as $b(z = 0) > 0$. In the top (bottom) panel of \autoref{fig:bias_contours}, we compare the posterior contours of a constant bias with those resulting from using a parameterisation where the bias evolves linearly (exponentially). In both cases, introducing more freedom to the bias model leads to a larger uncertainty in the bias, but the lower bounds on $A_\mathrm{ISW}$ are largely unaffected by the bias parameterisation. However, we can also observe that if the bias evolves more strongly with redshift, slightly larger values of $A_\mathrm{ISW}$ are likely, and overall, the evolving bias parameterisations favour to some degree higher values of $A_\mathrm{ISW}$, bringing its best-fitting value closer to unity (cf. \autoref{tab:bias_models_AISW_results}). For the most part though, the marginalised posterior distribution of $A_\mathrm{ISW}$ is robust under different bias parameterisations.

 \begin{table*}
    \centering
    \caption{Maximum posterior values of the galaxy bias $b_0$ at redshift $z=0$, its redshift evolution expressed as  $\mathrm{d}b/\mathrm{d}z$ or $\mathrm{d}\ln(b)/\mathrm{d}z$ and $A_\mathrm{ISW}$ from jointly analysing $\tilde{C}_\ell^\mathrm{gg}$ and $\tilde{C}_\ell^\mathrm{gT}$. The effective redshift $z_\mathrm{eff}$ is obtained as in \autoref{tab:b_evo_results}. We omit $\ell \leq 40$ when analysing the gg spectrum.}
    \label{tab:bias_models_AISW_results}
    \begin{tabular}{rcccccc}
        \hline
		Model & $b_0$ & $A_\mathrm{ISW}$ & $\frac{\mathrm{d}b}{\mathrm{d}z}$ & $\frac{\mathrm{d}\ln(b)}{\mathrm{d}z}$ & $z_\mathrm{eff}$ 
		& $b(z_\mathrm{eff})$\\ 
		\hline
		\multicolumn{5}{l}{SKADS}\\
		const. $b(z)$ & $3.248^{+0.068}_{-0.094}$ & $0.82^{+0.39}_{-0.33}$ & -- & -- & $1.56$ & $3.248$\\ 
		linear $b(z)$ & $2.60^{+0.78}_{-0.59}$ & $0.87^{+0.48}_{-0.31}$ & $0.73^{+0.73}_{-0.78}$ & -- & $1.78$ & $3.900$\\ 
		exp. $b(z)$ & $2.82^{+0.65}_{-0.59}$ & $0.89^{+0.43}_{-0.34}$ & -- & $0.13^{+0.24}_{-0.22}$ & $1.69$ & $3.513$\\ 
		\hline
		\multicolumn{5}{l}{T-RECS}\\
		const. $b(z)$ & $2.407^{+0.059}_{-0.061}$ & $0.86^{+0.32}_{-0.39}$ & -- & -- & $1.13$ & $2.407$\\ 
		linear $b(z)$ & $1.10^{+0.70}_{-0.45}$ & $0.96^{+0.41}_{-0.38}$ & $1.87^{+0.66}_{-0.79}$ & -- & $1.43$ & $3.774$\\ 
		exp. $b(z)$ & $1.56^{+0.32}_{-0.38}$ & $0.99^{+0.33}_{-0.42}$ & -- & $0.59\pm 0.24$ & $1.50$ & $3.852$\\ 
        \hline
    \end{tabular}
\end{table*}

Finally, we check for the impact of our ignorance on $n(z)$. We have fairly good knowledge of $n(z)$ for redshifts $z \lesssim 2$. The high-redshift tail of the distribution is one of the largest sources of uncertainty in the use of radio continuum surveys. This is reflected in the differences between the $n(z)$ estimated from SKADS and the one estimated from T-RECS, as plotted in \autoref{fig:nz}. We, therefore, repeat all the analyses done so far using the $n(z)$ from T-RECS and also list their results in \autoref{tab:bias_models_AISW_results}. Using the T-RECS $n(z)$, we obtain significantly lower values of the bias $b$ at redshift $z = 0$, but also significantly stronger redshift evolution, such that at the effective redshift $z_\mathrm{eff}$, when allowing for redshift evolution, the bias is roughly the same as when using the SKADS $n(z)$ (cf. \autoref{tab:b_AISW_results}). As can be seen in \autoref{fig:bias_models}, at redshift $z\sim 1$, also a constant bias with SKADS $n(z)$ agrees with the evolving parameterisations. Despite T-RECS's preference for strong bias evolution, we can see in \autoref{fig:nbz} that the product of $n(z)$ and the best-fitting $b(z)$ is generally unaffected by the choice of $n(z)$ and bias model. T-RECS favours a stronger localisation of objects below $z < 2$ and suppresses the high-redshift tail present in SKADS. The T-RECS analysis provides us with 68 per cent-credible intervals that are almost equal to the ones from the SKADS analysis, as can be seen by comparing red and blue whiskers in the top panel of \autoref{fig:aisw_boxplot}. Nevertheless, we obtain larger best-fitting values, with $A_\mathrm{ISW} = 0.99^{+0.33}_{-0.42}$ measured by assuming the T-RECS $n(z)$ and an exponentially evolving bias being the closest to one.

We are going to use the scatter among these different predictions to estimate the systematic uncertainty of our final $A_\mathrm{ISW}$ result.

\subsection{Comparison of Bias Measurements with Previous Models}
\label{sec:bias_results}

Before presenting a combined final result of our $A_\mathrm{ISW}$, we compare briefly our phenomenological bias results with previous results.

Most studies consider a bias model based on  N-body dark matter simulations \citep[e.g.][]{skads,2019MNRAS.482....2B}, in which the bias for each population of radio-galaxies is defined as belonging to a given halo mass $M_h$. To check the robustness of this approach, we have used the halo masses of each species of radio galaxy \citep[as specified in][]{2019MNRAS.482....2B}. For each galaxy type $i$ in the T-RECS medium sample:
\begin{equation}
    b_i(z)=\frac{\int{\de Mn_i(M,z)b_h(M,z)}}{\int{\de Mn_i(M,z)}}
    \label{eq:bias_colossus}
\end{equation}
where $n_i(M,z)$ is the halo mass function for galaxies of population type $i$ in the redshift bin $(z,z+\de z)$ and $b_h(M,z)$ is the halo bias, which we estimate using Colossus suite \citep{Diemer2018} and the halo model from \citet{Tinker2010}. The total bias is then computed using \autoref{eqn:total_bias}.

\begin{figure}
    \centering
    \includegraphics[width=\columnwidth]{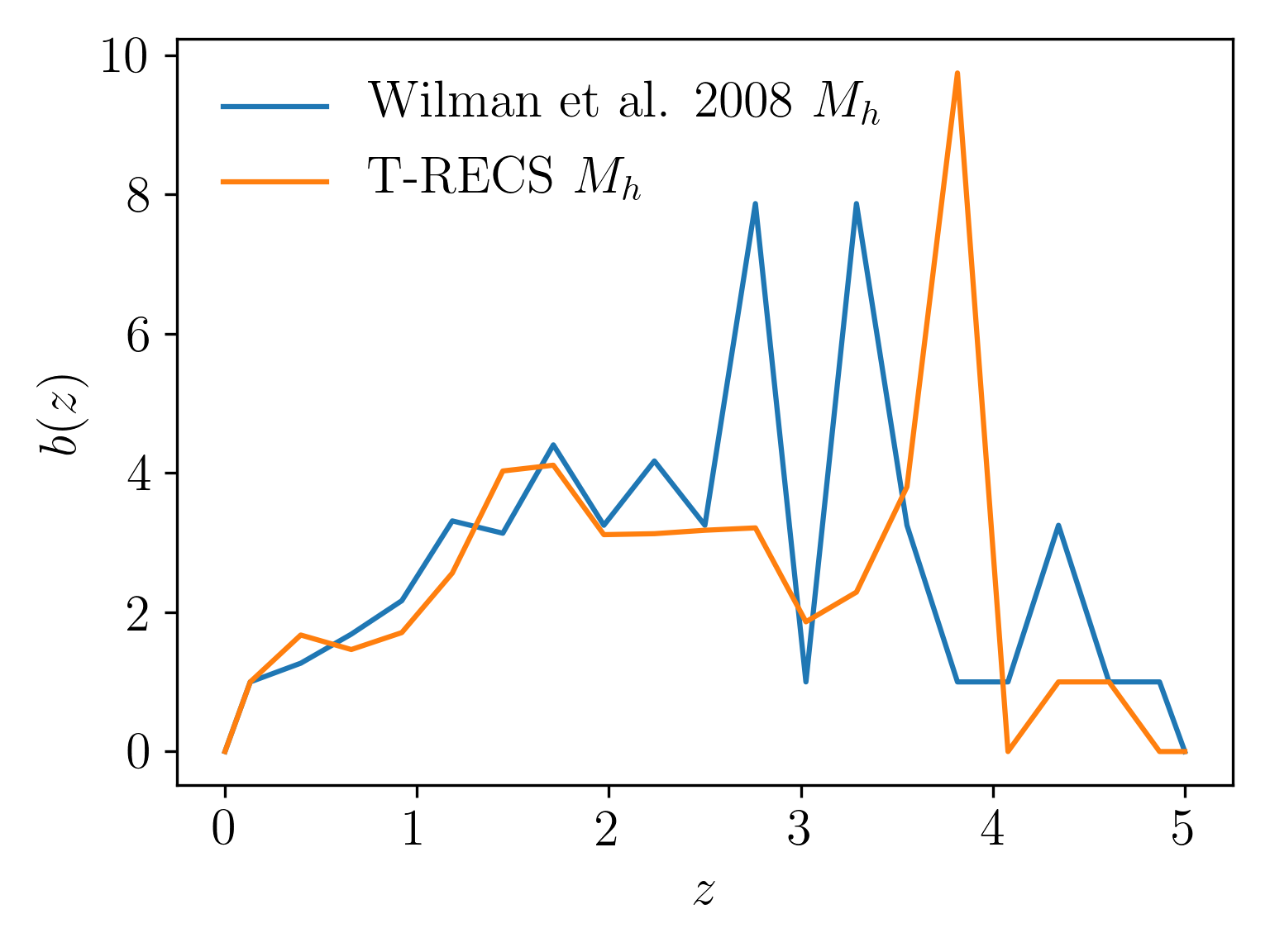}
    \caption{Ensemble bias for the T-RECS medium simulation for a flux cut of 4 mJy for an $888\, \mathrm{MHz}$ catalogue. In this case, we fix the redshift at which the bias is evaluated for high redshifts, depending on the population.}
    \label{fig:bias_trecs1}
\end{figure}

\begin{figure}
    \centering
    \includegraphics[width=\columnwidth]{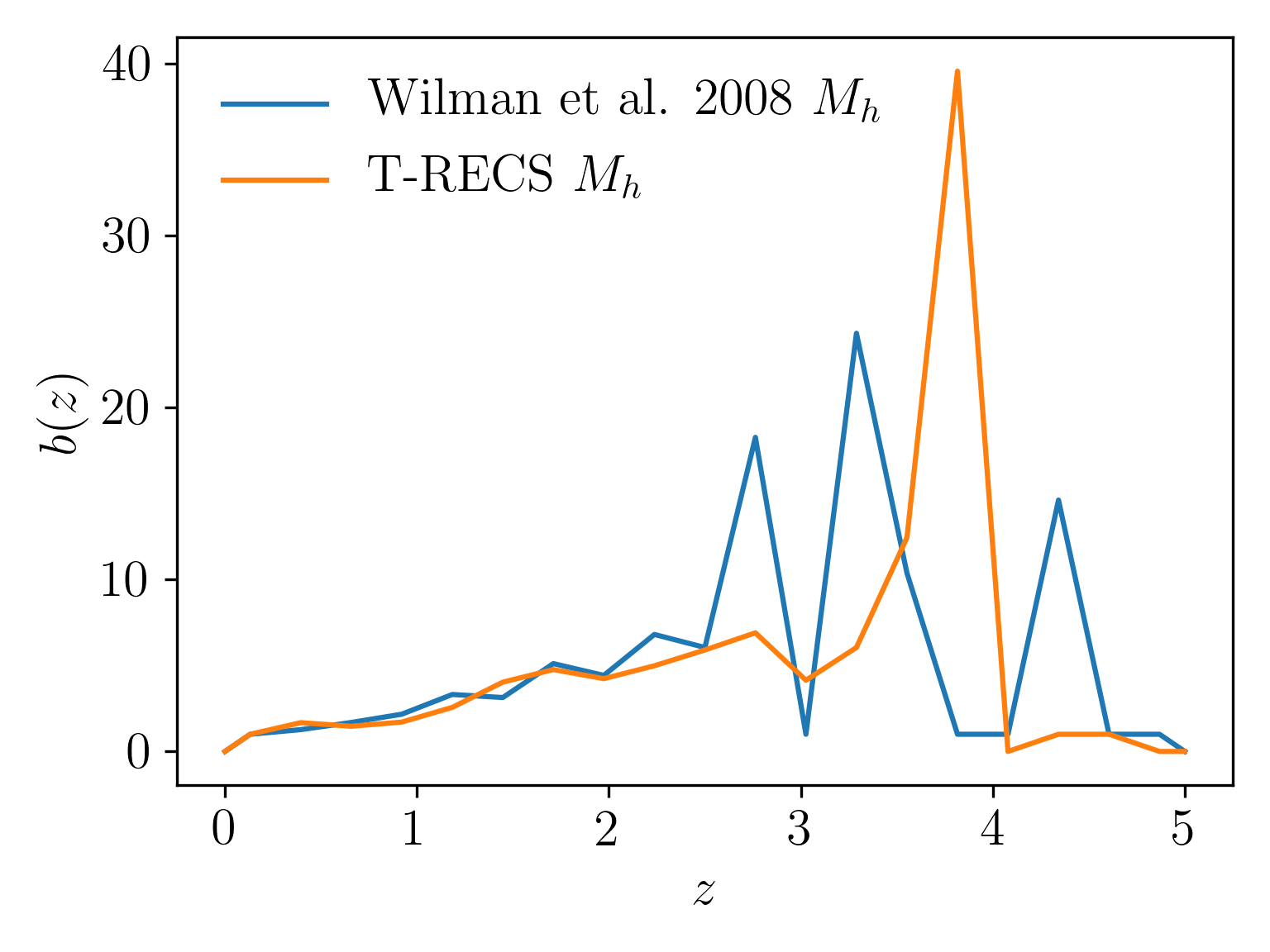}
    \caption{Same as \autoref{fig:bias_trecs1} but without redshift cut when evaluating the halo bias in \autoref{eq:bias_colossus}.}
    \label{fig:bias_trecs2}
\end{figure}

In \autoref{fig:bias_trecs1}, we show the T-RECS  total bias $b(z)$ redshift evolution when using a flux cut of 4 mJy at $888\, \mathrm{MHz}$, using the approach above. As proposed in \cite{skads}, for high redshifts, we evaluate \autoref{eq:bias_colossus} at a fixed redshift ($z=1.5$ or $z=3$ depending on the population).  We compare this result with the \cite{skads} approach by using the SKADS simulation, in which we evaluate the halo bias at the corresponding halo mass value proposed for the SKADS simulation for each population type. We see that the estimated bias is similar for both approaches.  We also see that if we mix information from both simulations, we obtain a wrong result. The scattered and noisy behaviour at larger redshifts is due to the small number of halos that remain after all the applied cuts. In \autoref{fig:bias_trecs2}, we show the same $b(z)$ but without applying any redshift cut off on the bias evaluation. We see that the bias grows, both for SKADS and T-RECS to extremely high values due to the FRII galaxy population.


As the T-RECS $n(z)$ model predicts only a few objects at high redshifts, we still see agreement between the N-body result and our measurements of the T-RECS $n(z)b(z)$ high-redshift tail (cf. \autoref{fig:nbz}). Surprisingly, the N-body based T-RECS model underpredicts $n(z)b(z)$ at redshifts $z\lesssim 1$. This might be due to the SFG abundance being underpredicted and needs further investigation in the preparation for EMU.

We also added the parameterised form of the bias as given as part of the \citet{skads} analysis to \autoref{fig:nbz}. Similar as in the T-RECS case, we measure a lower $n(z)b(z)$ at $z\lesssim 1$ with RACS as the N-body simulations suggest, thus, SKADS might also predict less SFGs as there are in reality. Furthermore, we see that $n(z)b(z)$ peaks at a much higher redshift than all best-fitting models. The peak is where we expect FRIIs to dominate the sample. We deduce from this that the abstruse FRII bias has so far been over-estimated. Another possible explanation for the mismatch in \autoref{fig:nbz} could be that FRIIs start to dominate at higher redshifts than previously thought. We leave a detailed examination of this issue for future work.

\subsection{Combining Different Predictions}
\label{sec:baccus}

\begin{figure}
    \centering
    \includegraphics[width = \columnwidth]{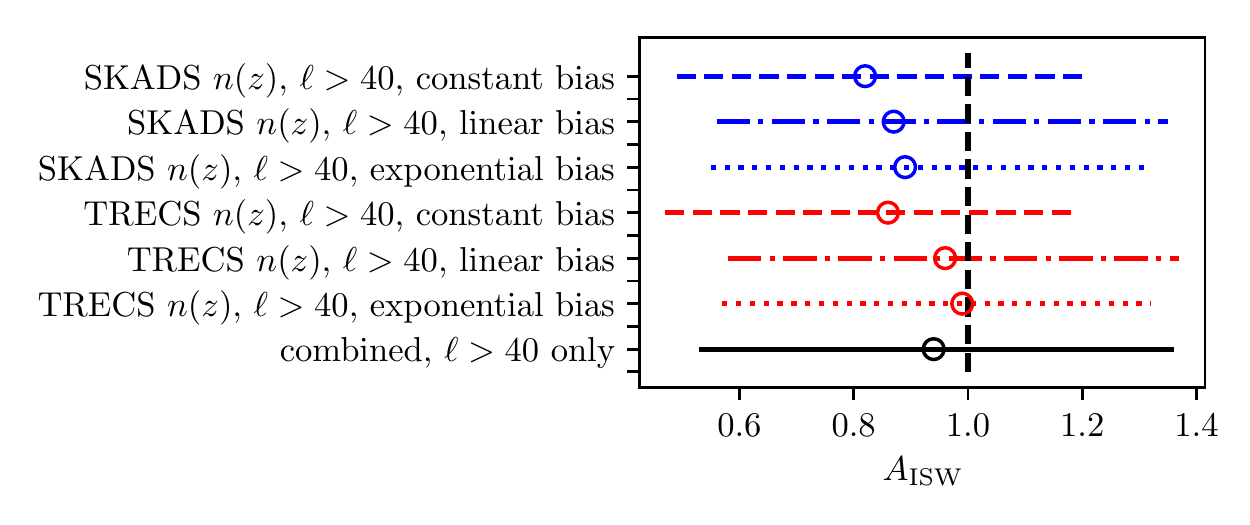}
    \includegraphics[width = \columnwidth]{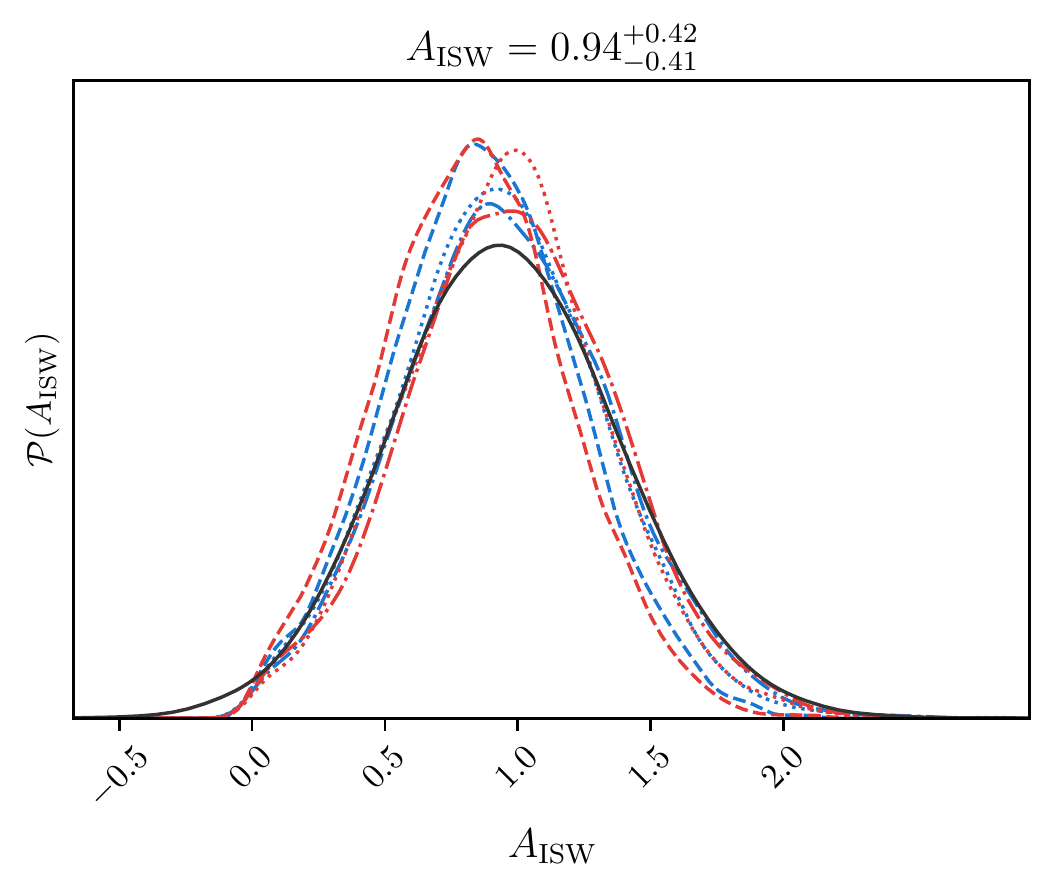}
    \caption{{\it Top:} Boxplot summarising the $A_\mathrm{ISW}$ results obtained using different $n(z)$ and bias parameterisations, leaving out $\tilde{C}_\ell^{\mathrm{gg}}$ at $\ell \leq 40$. Potential systematic biases are lessened when the different $n(z)$ and bias models are combined using the BACCUS approach (black line). {\it Bottom:} The corresponding marginalised posterior distributions on $A_\mathrm{ISW}$.}
    \label{fig:aisw_boxplot}
    \label{fig:AISW_posterior}
\end{figure}

In \autoref{fig:aisw_boxplot}, we show the best-fitting values and marginalised posterior distributions for $A_\mathrm{ISW}$ obtained using different bias models and $n(z)$ distributions (for an extended version of this plot including $\tilde{C}_\ell^\mathrm{gg}$ data at $\ell \leq 40$, we refer to Appendix \ref{sec:isw_with_large_scales}). Given their scatter, we can estimate the overall uncertainty $A_\mathrm{ISW}$ including the uncertainty due to unknown systematics. To avoid expectation bias, we follow the BACCUS (BAyesian Conservative Constraints and Unknown Systematics) approach of \citet{LuisBernal:2018drn}, where we assume that unknown systematic effects have biased all of our measurements $A_\mathrm{ISW}^{(i)}$ by an unknown $\Delta A_\mathrm{ISW}^{(i)}$ and that they also degrade the variance $\sigma_i^2/\zeta_i$ by a factor of $\zeta_i$. As all our $A_\mathrm{ISW}$ results are single-peaked but asymmetric, we summarise their marginalised posterior distributions in terms of Variable Gaussians, i.e.\ as a Normal distribution whose scale parameter $\sigma\left(A_\mathrm{ISW}\right)$ depends on the value of $A_\mathrm{ISW}$ in the exponential \citep[while keeping the log-determinant fixed, ][]{Bartlett:1953}:
\begin{equation}
    -2\ln\mathcal{P}\left(\left.A_\mathrm{ISW}\right\vert \sigma_i, A_\mathrm{ISW}^{(i)}\right) = \left(\frac{A_\mathrm{ISW} - A_\mathrm{ISW}^{(i)}}{\sigma_i\left(A_\mathrm{ISW}\right)}\right)^2 + \text{const.}
\end{equation}
Assuming a linear relationship and imposing that the asymmetric errors $\sigma_i^+$ and $\sigma_i^-$ define the full half-width maximum (FHWM), one finds \citep{Barlow:2004wg}
\begin{equation}
    \sigma_i\left(A_\mathrm{ISW}\right)=\frac{2\sigma_i^+\sigma_i^-}{\sigma_i^++\sigma_i^-} + \frac{\sigma_i^+ - \sigma_i^-}{\sigma_i^+ + \sigma_i^-} \left(A_\mathrm{ISW} - A_\mathrm{ISW}^{(i)}\right).
\end{equation}
As $A_\mathrm{ISW}$ is also statistically independent from the galaxy bias parameters, introducing the systematic bias $\Delta A_\mathrm{ISW}^{(i)}$ and variance degradation parameters $\zeta_i$, we write the log-likelihood of each measurement (dropping constant terms) as 
\begin{align}
    2\ln\mathcal{P}&\left(\left.A_\mathrm{ISW}, \zeta_i, \Delta A_\mathrm{ISW}^{(i)}\right\vert \sigma_i, A_\mathrm{ISW}^{(i)}\right) \nonumber\\&= \ln\left(\zeta_i\right) - \zeta_i\left(\frac{A_\mathrm{ISW} + \Delta A_\mathrm{ISW}^{(i)} - A_\mathrm{ISW}^{(i)}}{\sigma_i\left(A_\mathrm{ISW}\right)}\right)^2.
\end{align}
Since measurements $(i)$ come from the same data, we cannot just add up the individual log-likelihoods. Instead, we consider the average log-likelihood
\begin{align}
    \ln\mathcal{P}&\left(\left.A_\mathrm{ISW}, \bzeta, \bm{\Delta A}_\mathrm{ISW}\right\vert \bsigma, \bm{A}_\mathrm{ISW}\right)\nonumber\\
    &=\frac{1}{6}\sum_{i=0}^{6}\ln\mathcal{P}\left(\left.A_\mathrm{ISW}, \zeta_i, \Delta A_\mathrm{ISW}^{(i)}\right\vert \sigma_i, A_\mathrm{ISW}^{(i)}\right).
    \label{eq:baccus_lik}
\end{align}
We follow \citet{LuisBernal:2018drn} in our choice of priors on the scaling parameters $\bsigma$ and the systematic bias shifts $\bm{\Delta A}_\mathrm{ISW}$. Thus, we assume that we have estimated the size of our statistical errors correctly, and, as a consequence, that the prior on the scaling parameters is \citep*{Hobson:2002zf}
\begin{equation}
    \mathcal{P}\left(\zeta_i\right) \propto \begin{cases}\exp\left(-\zeta_i\right)&\text{ if }\zeta_i >0,\\0&\text{ else.}\end{cases}
\end{equation}
We also choose a zero-centred Gaussian prior on $\Delta A_\mathrm{ISW}^{(i)}$ with width $\sigma_i$. 

We display at the bottom of \autoref{fig:AISW_posterior} the posterior on $A_\mathrm{ISW}$ after marginalising over $\bzeta$ and $\bm{\Delta A}_\mathrm{ISW}$. We find a best-fitting value of $A_\mathrm{ISW} = 0.94^{+0.42}_{-0.41}$, thus $2.3\sigma$ away from $A_\mathrm{ISW} = 0$. Allowing for both statistical and systematic uncertainties in this conservative approach, we obtain a probability for a positive $A_\mathrm{ISW}$ of 98.9 per cent.

\section{Summary}
\label{sec:summary}

\begin{itemize}
    \item We have measured the angular power spectrum of the radio continuum sources detected above a $4\;\mathrm{mJy}$ flux density limit by the Rapid ASKAP Continuum Survey at $888\, \mathrm{MHz}$, in auto-correlation and also in cross-correlation with temperature maps of the cosmic microwave background from the \textit{Planck} mission.
    \item We constructed estimates of the variance of the angular power spectra, using the purely analytic prediction from theory, jack-knife resampling of the catalogue data, and two methods that use simulation of mock RACS catalogues using the Full-sky Lognormal Astro-fields Simulation Kit (FLASK) (sample covariance of the mocks, and a graphical lasso estimator learning sparse covariance matrices from simulations). All of these gave roughly consistent results, with the sample-covariance and jack-knife approaches predicting more off-diagonal covariance compared to the others.
    \item We have tested four different bias parameterisations, the goodness of fit is almost indistinguishable among them, making it impossible to pick only one of them. We have found that the product of the best-fitting biases $b(z)$ and the redshift distributions of sources per steradian $n(z)$ has lower values at $z \ga 2$ than what we had predicted from SKADS and T-RECS simulations, hinting towards the assumed FRII bias value being too large.
    \item We have found that the angular auto-power spectrum of RACS galaxies is consistent with the prediction from $\Lambda$CDM, except on large scales, $\ell \leq 40$, where we detect an excess which we believe is due to systematics.
    \item We have split the RACS catalogue into different regions and measured the angular power spectrum, to test for systematic causes for the excess. We have found a tentative trend showing that the large-scale excess is more pronounced for regions with declination closer to the equator. However, the error on the power spectrum estimated from mocks also increases towards the equator, thus, the measured power is consistent with the $\Lambda$CDM prediction in almost all declination strips. In comparison, we see no such trend with Right Ascension. This strongly implies that the excess is due to a systematic effect associated with the noise or some other observational effect.
    \item We have detected a cross-correlation between the galaxy distribution and the distribution of hot and cold spots in the cosmic microwave background. This cross-correlation has been measured through the angular power cross-spectrum $\mathcal{C}_{\ell}^{gT}$, and is significant at 2.8$\sigma$ relative to the null hypothesis of no cross-correlation.
    \item We have found that when fitting the data from both the auto- and cross-correlation, the fit is consistent with the $\Lambda$CDM prediction (when the $\ell \leq 40$ data is removed from the auto-power spectrum).
    \item We have parameterised the amplitude of the cross-correlation signal $A_{\mathrm{ISW}}$. We find that when combining the angular auto- and cross-power spectra, and assuming an $n(z)$ from SKADS and a constant bias model, that $A_{\mathrm{ISW}}=0.82^{+0.39}_{-0.33}$. These constraints are not very sensitive to the choice of the number count model or bias model.
    \item When using the BACCUS approach to marginalise over different assumptions on the bias and number count model, allowing for unknown systematic biases and for possible posterior widening due to unknown systematic effects, we have found $A_\mathrm{ISW} = 0.94^{+0.42}_{-0.41}$, corresponding to a $2.3\sigma$ or 98.9 per cent detection of the ISW effect and, hence, of dark energy.

This analysis has demonstrated that a few weeks on-source time of ASKAP observations provide data for meaningful cosmological analyses, while identifying what points have to be addressed in the analysis pipeline to reap the full potential of the upcoming EMU survey. The cosmological utility of the clustering statistics of radio continuum galaxies can only improve through the pathfinder era, to reach maturity with the SKA Observatory.
\end{itemize}

\section*{Acknowledgements}

BBK, DP, and FQ are supported by the project \begin{CJK}{UTF8}{mj}우주거대구조를 이용한 암흑우주 연구\end{CJK} (``Understanding Dark Universe Using Large Scale Structure of the Universe''), funded by the Ministry of Science. SC acknowledges support from the `Departments of Excellence 2018-2022' Grant (L.\ 232/2016) awarded by the Italian Ministry of University and Research (\textsc{mur}). JA has received funding from the European Union's Horizon 2020 research and innovation programme under grant agreement No. 776247 EWC. CLH is supported by a Leverhulme Trust Early Career Research Fellowship.

This work was supported by the high performance computing clusters Seondeok at the Korea Astronomy and Space Science Institute. This research made substantial use of \texttt{Astropy},\footnote{\url{http://www.astropy.org}} a community-developed core Python package for Astronomy \citep{astropy:2013, astropy:2018}, \texttt{CAMB} \citep{Lewis:1999bs,Howlett:2012mh}, the \texttt{ChainConsumer} package \citep{samuel_hinton_2020_4280904}, the \texttt{flask} package \citep{Xavier:2016elr}, the \texttt{HEALPix} and \texttt{healpy} packages \citep{2005ApJ...622..759G,Zonca2019}, \texttt{NaMaster} \citep{2002ApJ...567....2H,2019MNRAS.484.4127A}, the \texttt{Numpy} package \citep{book}, the \texttt{Scikit-learn} package \citep{scikit-learn}, the \texttt{Scipy} package \citep{2019arXiv190710121V}, \texttt{Matplotlib} \citep{Hunter:2007} and \texttt{zeus} \citep{karamanis2021zeus}.

The Australian SKA Pathfinder is part of the Australia Telescope National Facility (\hyperlink{https://ror.org/05qajvd42}{https://ror.org/05qajvd42}) which is managed by CSIRO. Operation of ASKAP is funded by the Australian Government with support from the National Collaborative Research Infrastructure Strategy. ASKAP uses the resources of the Pawsey Supercomputing Centre. Establishment of ASKAP, the Murchison Radio-astronomy Observatory and the Pawsey Supercomputing Centre are initiatives of the Australian Government, with support from the Government of Western Australia and the Science and Industry Endowment Fund. We acknowledge the Wajarri Yamatji people as the traditional owners of the Observatory site.

\section*{Data Availability}

The RACS radio continuum stokes I source catalogue used in this analysis was generated from data available from the CSIRO ASKAP Science Data Archive (CASDA). The angular power spectra and covariance matrices that were measured are made available via a GitHub repository, which can be found at \hyperlink{https://github.com/racs-cosmology/isw}{https://github.com/racs-cosmology/isw}. The MCMC chains generated for the analysis are available on request to the authors.



\bibliographystyle{mnras}
\bibliography{references} 




\appendix

\section{Is there an ``Axis of Evil''?}
\label{sec:lsssyst}

\begin{figure}
    \centering
    \includegraphics[width = \columnwidth]{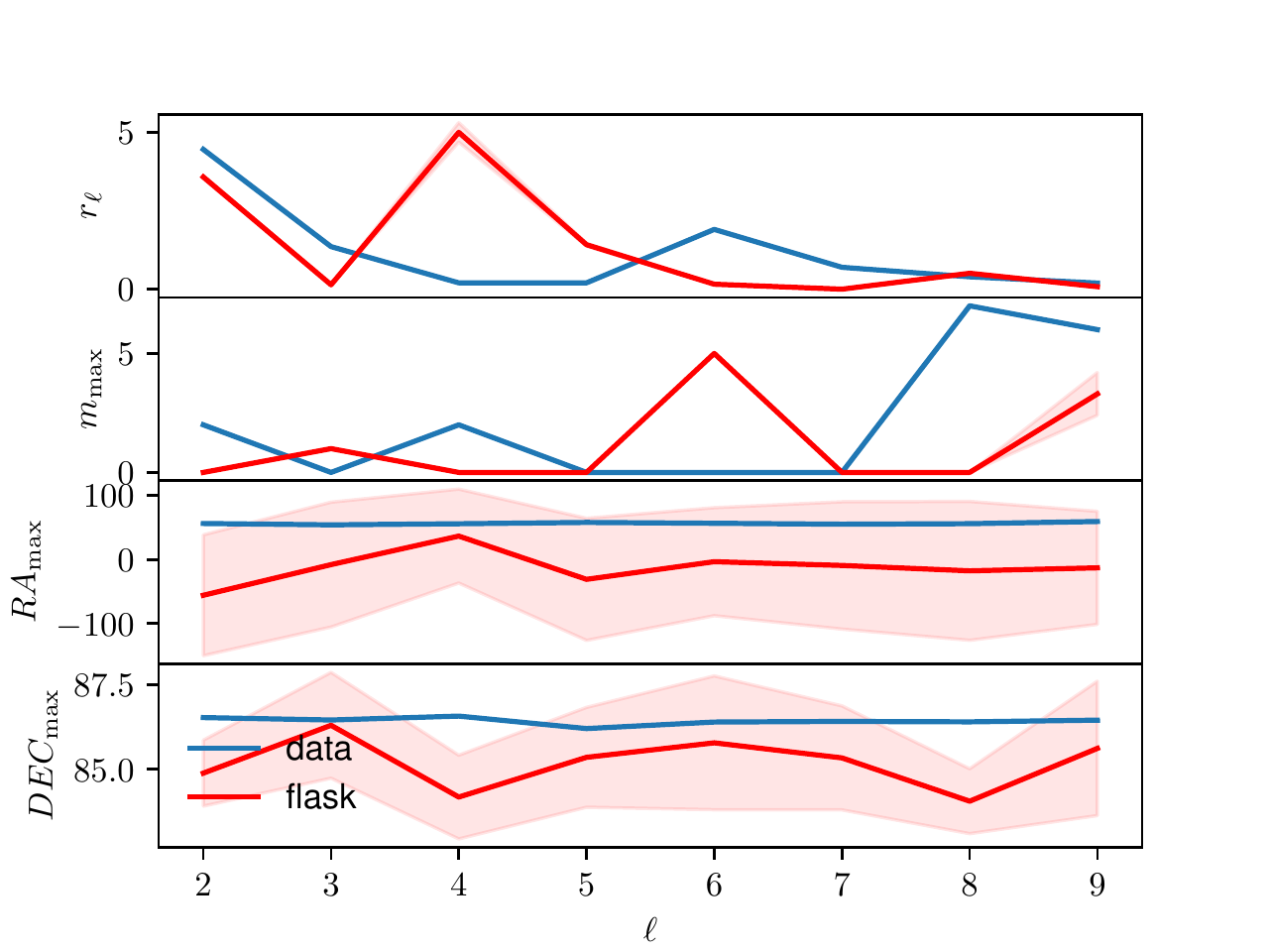}
    \caption{The ratio $r_\ell$ of power absorbed by the maximum mode with ``shape'' $m_\mathrm{max}$ in direction $\bm{n}_\mathrm{max} = (\mathrm{Dec}_\mathrm{max}, \mathrm{RA}_\mathrm{max})$. The blue line has been obtained from the data. The magenta line shows the mean of 10 flask realisations, whereas the shaded region displays the 1-$\sigma$ region estimated from the same 10 realisations.}
    \label{fig:axis_of_evil}
\end{figure}

We have omitted multipoles at $\ell \leq 40$ in our $C_\ell^\mathrm{gg}$ analyses due to an excess in the power spectrum that cannot be described by our model.
Whilst we believe this is due to systematics, one possible cause of this excess power might also be a large anisotropy in the distribution of continuum galaxies. It would be possible to test this explanation in harmonic space, by conducting a similar analysis to that which detected a preferred axis of the cosmic microwave background anisotropy \citep{2005PhRvL..95g1301L}. We consider the ratio 
\begin{equation}
 r_\ell = \max_{m\bm{n}}\frac{C_{\ell m}}{(2\ell + 1)C_\ell}
\end{equation}
of power absorbed by the maximum mode with ``shape'' $m_\mathrm{max}$ in direction $\bm{n}_\mathrm{max}$,
where 
\begin{equation}
    C_{\ell m} = \begin{cases} \left\vert a_{\ell m}\right\vert^2, & \text{for }\ell = 0\\
   2 \left\vert a_{\ell m}\right\vert^2, & \text{else,} \end{cases}
\end{equation}
and which we plot in \autoref{fig:axis_of_evil}. We compute $r_\ell$ for the data and for 10 flask realisations. Surprisingly, the direction $\bm{n}_\ell$ of the data is within the 1-$\sigma$ bounds of the flask directions, but the data and flask $r_\ell$ are discrepant. In any case, the data $r_\ell$ have the same magnitude as the ones we see in the flask realisations. The fact that the flask realisations show almost no scatter at most multipoles suggests that this is mostly driven by the mask. This is further supported by the "axis of evil" pointing towards a direction close to the North Pole, around which the RACS mask is almost symmetric (cf. \autoref{fig:racs_data}).
 
\section{Measuring ISW excluding large-scale multipoles in both the galaxy-galaxy auto power spectrum and galaxy-temperature cross power spectrum}
\label{sec:isw_without_large_scales}

\begin{figure}
    \centering
    \includegraphics[width = \columnwidth]{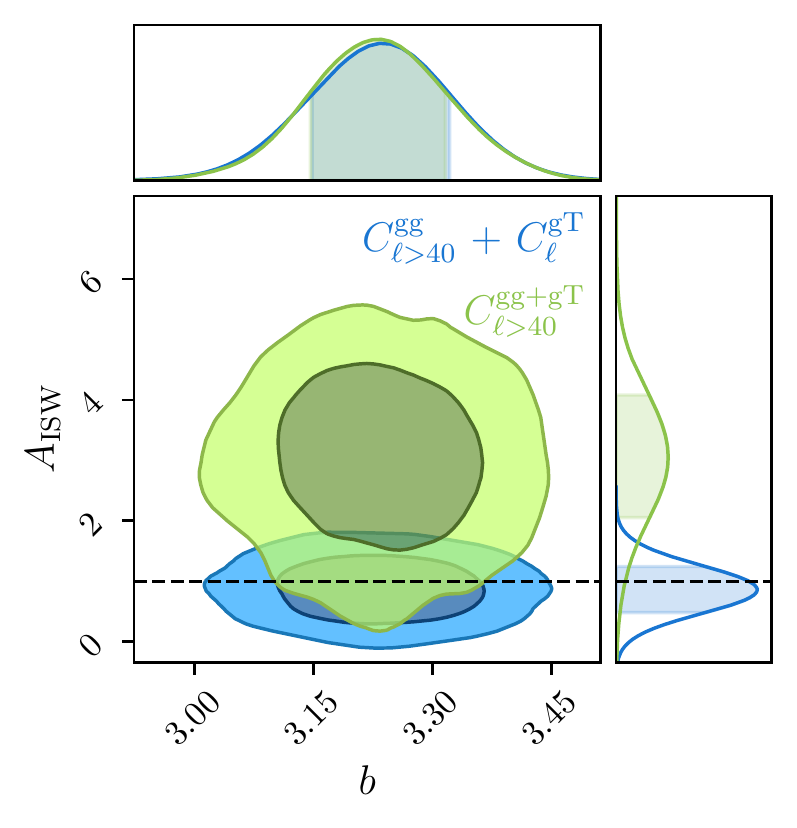}
    \caption{Comparison of the $b$-$A_\mathrm{ISW}$ contours from a combined $C_\ell^\mathrm{gg}$ and $C_\ell^\mathrm{gT}$ fit. In both cases, we omit large-scale multipoles ($\ell < 40$) in the galaxy-galaxy auto power spectrum. For the blue contour, we use the full available multipole range in the galaxy-temperature cross power spectrum, whereas for the green contour, we also omit $\ell < 40$ in the gT spectrum. The dashed line marks the expectation of $A_\mathrm{ISW} = 1$.}
    \label{fig:RACS_b_AISW_cut_in_gT_cornerplot}
\end{figure}

In \autoref{sec:gTspectrum}, we have argued that, even though we ignore large-scale $C_\ell^\mathrm{gg}$-multipoles, we can still trust large-scale $C_\ell^\mathrm{gT}$-multipoles. We present in \autoref{fig:RACS_b_AISW_cut_in_gT_cornerplot} a combined measurement of a constant bias $b$ and $A_\mathrm{ISW}$ where we omit $\ell < 40$ in both $C_\ell^\mathrm{gg}$ and $C_\ell^\mathrm{gT}$. Without the first two $\ell$-bins, \autoref{fig:cl_RACS} shows that the ISW signal is only distinguishable from the null hypothesis of no galaxy-temperature correlation in the third and fourth $\ell$-bin. We can also see there, that the third $\ell$-bin has a larger value of $C_\ell^\mathrm{gT}$ than expected from neighbouring values. Without the first two $\ell$-bins, we therefore see an increased value of $A_\mathrm{ISW} = 3.09^{+0.99}_{-1.03}$. Thus, using the full $\ell$-range is actually more conservative since cutting out large-scale multipoles pushes the significance of the ISW detection up to $3\;\sigma$. As can also be seen in \autoref{fig:RACS_b_AISW_cut_in_gT_cornerplot}, the bias is unaffected by the large-scale gT power.

\section{ISW constraints including large-scale multipoles}
\label{sec:isw_with_large_scales}

\begin{figure}
    \centering
    \includegraphics[width = \columnwidth]{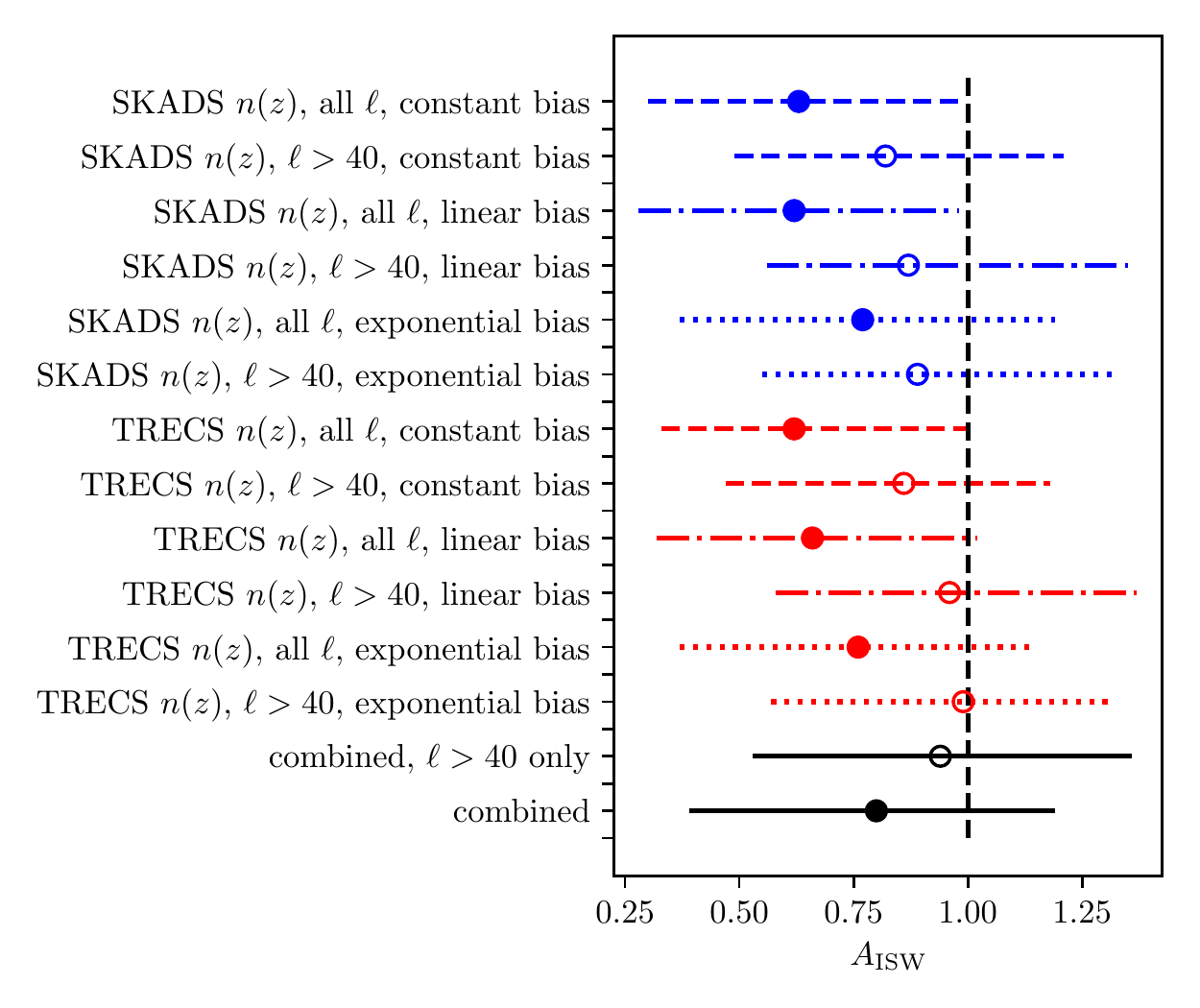}
    \caption{Boxplot summarising the $A_\mathrm{ISW}$ results obtained using different $n(z)$, $\ell$ ranges and bias parameterisations. Including all of the data, including the $\tilde{C}_\ell^{\mathrm{gg}}$ for $\ell \leq 40$ (the solid circles) lowers the mean value of $A_\mathrm{ISW}$ by approximately 0.5$\sigma$, in comparison to estimates where this data is left out (empty circle). This potential systematic bias is lessened when the different models are combined using the BACCUS approach (black line).}
    \label{fig:aisw_boxplot_all}
\end{figure}

In this appendix, we present the results using the full available multipole range also in $C_\ell^\mathrm{gg}$. In line with \autoref{fig:RACS_b_AISW_cornerplot}, for all bias and $n(z)$ models, larger values of the galaxy bias $b(z)$ are favoured when including multipoles at $\ell \leq 40$ as the bias is the only parameter we vary in our $C_\ell^\mathrm{gg}$ model. As the amplitude of $b(z)$ is degenerate with $A_\mathrm{ISW}$ in $C_\ell^\mathrm{gT}$, the full-range analysis hence supports lower values of $A_\mathrm{ISW}$, which we present in \autoref{fig:aisw_boxplot_all}. 

As can be seen there, even when we consider the $\ell$-range where our $C_\ell^\mathrm{gg}$ model breaks down, there is no set of $b(z)$ and $n(z)$ model where the data is consistent with $A_\mathrm{ISW} = 0$. This result is not unexpected, given that we believe the $\ell \leq 40$ auto-power spectrum excess to be a systematic, and so uncorrelated with the CMB photon distribution.

\begin{figure}
    \centering
    \includegraphics[width = \columnwidth]{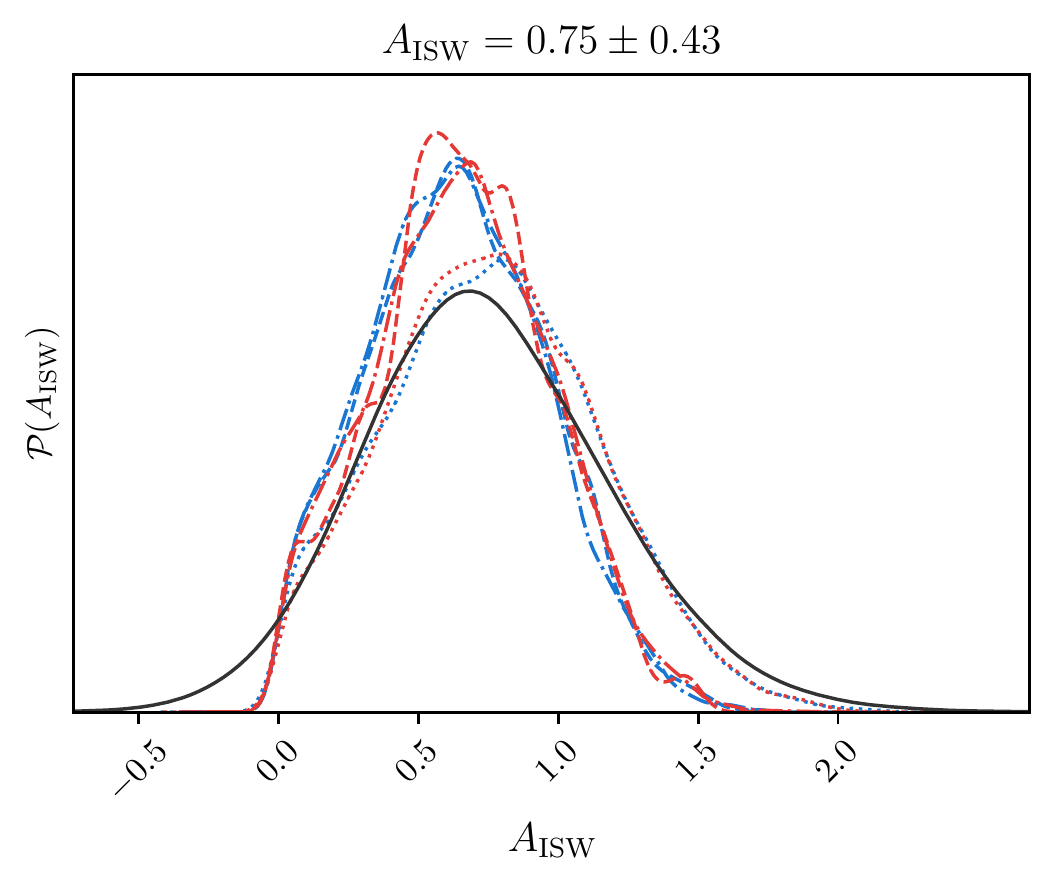}
    \caption{The posterior on $A_\mathrm{ISW}$ after combining the measurements presented in \autoref{fig:aisw_boxplot_all} in a BACCUS-like \citep{LuisBernal:2018drn} fashion.}
    \label{fig:AISW_posterior_all}
\end{figure}

We repeat the BACCUS-like analysis described in \autoref{sec:baccus}. The combined posterior distribution of $A_\mathrm{ISW}$ is graphed in \autoref{fig:AISW_posterior_all}. We measure $A_\mathrm{ISW} = 0.75\pm 0.43$, thus even when we include the $\ell \leq 40$-range, the data is in favour of the ISW effect with $1.7 \sigma$. It has to be noted, however, that the Variable Gaussian distribution does not summarise the individual $A_\mathrm{ISW}$ posteriors well and that the true low-$A_\mathrm{ISW}$ tails are less pronounced than those of the Gaussian approximations (cf. \autoref{fig:AISW_posterior_all}). Our estimate of $1.7 \sigma$ is therefore conservative.

\bsp	
\label{lastpage}
\end{document}